\documentclass[twocolumn,secnumarabic,amssymb, nobibnotes, aps, prx,superscriptaddress]{revtex4-1}

\usepackage[paperwidth=210mm,paperheight=297mm,centering,hmargin=2cm,vmargin=2.5cm]{geometry}
\usepackage{amsmath} % Advanced math typesetting
\usepackage[english]{babel} % Change hyphenation rules
\usepackage{graphicx} % Add pictures to your document
\usepackage{listings} % Source code formatting and highlighting
\usepackage{soul}
\usepackage{color}
\usepackage{lipsum}
\usepackage{float}

\begin{document}

\title{Beyond uniform screening: electrostatic heterogeneity dictates solution structure of complex macromolecules}

\author{Fabrizio Camerin}
\email[Corresponding author:]{ fabrizio.camerin@chem.lu.se}
\affiliation{Division of Physical Chemistry, Department of Chemistry, Lund University, Lund, Sweden}

\author{Marco Polimeni}
\affiliation{Department of Pharmacy, Drug Delivery and Biophysics of Biopharmaceuticals, University of Copenhagen, Copenhagen, Denmark}

\author{Letizia Tavagnacco}
\affiliation{CNR Institute of Complex Systems, Uos Sapienza, Piazzale Aldo Moro 2, 00185 Roma, Italy}
\affiliation{Department of Physics, Sapienza University of Rome, Piazzale Aldo Moro 2, 00185 Roma, Italy}

\author{Jeffrey C. Everts}
\affiliation{Institute of Theoretical Physics, Faculty of Physics, University of Warsaw, Poland}
\affiliation{Institute of Physical Chemistry, Polish Academy of Sciences, Poland}

\author{Szilard Sáringer}
\affiliation{Division of Physical Chemistry, Department of Chemistry, Lund University, Lund, Sweden}

\author{Alessandro Gulotta}
\affiliation{Division of Physical Chemistry, Department of Chemistry, Lund University, Lund, Sweden}

\author{Nicholas Skar-Gislinge}
\affiliation{Division of Physical Chemistry, Department of Chemistry, Lund University, Lund, Sweden}

\author{Anna Stradner}
\affiliation{Division of Physical Chemistry, Department of Chemistry, Lund University, Lund, Sweden}
\affiliation{LINXS Institute of Advanced Neutron and X-ray Science, Lund University, Lund, Sweden}

\author{Emanuela Zaccarelli}
\email[Corresponding author:]{ emanuela.zaccarelli@cnr.it}
\affiliation{CNR Institute of Complex Systems, Uos Sapienza, Piazzale Aldo Moro 2, 00185 Roma, Italy}
\affiliation{Department of Physics, Sapienza University of Rome, Piazzale Aldo Moro 2, 00185 Roma, Italy}

\author{Peter Schurtenberger}
\email[Corresponding author:]{ peter.schurtenberger@chem.lu.se}
\affiliation{Division of Physical Chemistry, Department of Chemistry, Lund University, Lund, Sweden}
\affiliation{LINXS Institute of Advanced Neutron and X-ray Science, Lund University, Lund, Sweden}

\date{\today}

\begin{abstract}
\noindent

The complexity of biomolecular interactions necessitates advanced methodologies to accurately capture their behavior in solution. In this work, we focus on monoclonal antibodies and adopt a multi-scale coarse-graining strategy for their modeling, with particular emphasis on the role of electrostatic interactions. 
Using scattering experiments, theoretical analysis, and large-scale computer simulations, we explicitly compare two selected case studies—markedly different in their charge distributions. Through mutually corroborating lines of evidence, we demonstrate that conventional approaches relying on electrostatic screening and implicit charge representations fail to capture the structural and thermodynamic properties of antibody solutions when strong charge heterogeneity is present, even at a moderate (amino acid) level of coarse-graining.
These findings highlight the importance of a correct treatment of electrostatic interactions and ion screening for heterogeneously- and oppositely-charged colloidal and protein systems. Such considerations are essential to move beyond descriptive models towards a truly predictive framework, with direct implications for the formulation of therapeutics and the treatment of other complex soft-matter systems.

\end{abstract}

\maketitle

\section{Introduction}

One of the enduring goals of soft matter and biomolecular science is to connect molecular-scale interactions to emergent macroscopic properties~\cite{Cruz2025, Barrat2024}. Whether in the context of protein assembly or the design of functional colloids, the recurrent challenge is to build models that are minimal and tractable, yet sufficiently detailed to capture the essential physics and retain predictive power~\cite{noid2013perspective,belloni2000colloidal,likos2001effective, stradner2020potential}. Finding this balance is especially challenging when interactions act over long distances, are directional, and change with the surroundings.

Electrostatic interactions exemplify this difficulty. Unlike short-ranged forces, they couple local molecular features to collective behavior over mesoscopic length scales. For decades, coarse-grained models have relied on mean-field approximations of electrostatics, most notably the screened Coulomb or Yukawa potential derived from the linearized Poisson–Boltzmann equation~\cite{derjaguin1993theory}. This approach has been highly successful for suspensions of nearly uniformly charged colloids, enabling theoretical descriptions and simulations that capture their structure and phase behavior~\cite{trefalt2017forces,Naegele1996}. 
However, several works have shown that Yukawa interactions fail in more complex or limiting situations. In strongly deionized suspensions, for high salt concentrations or for strongly charged particles, the Yukawa description breaks down and ion correlations become dominant~\cite{torres2008breakdown,reinertsen2024reexpansion,wong2010electrostatics}. Like-charge attraction and long-ranged ordering phenomena, observed in both simulations and experiments, cannot be captured within a linearized mean-field framework~\cite{li2017strong,naderi2020self}. 
More recently, the phenomenon of underscreening in concentrated electrolytes has highlighted how real systems present screening lengths much longer than Debye theory predicts due to structural correlations~\cite{krucker2025potential,jager2023screening}. Taken together, these findings show that in most cases Yukawa models can, at best, offer only a partial description of the systems under investigation.

Throughout the years, numerous approaches have also been proposed to go beyond these shortcomings, ranging from charge renormalization schemes to nonlinear Poisson–Boltzmann treatments~\cite{belloni1998ionic,alexander1984charge,trizac2002simple,jansen2025surface}. Yet, each comes with limitations. For instance, some are tied to specific geometries, while others struggle to simultaneously capture local molecular-scale electrostatics and the mesoscopic behavior that experiments typically probe in complex environments~\cite{boon2010screening,lovsdorfer2013symmetry,obolensky2021rigorous}. Moreover, to our knowledge, a direct, one-to-one comparison between these theoretical descriptions and experimentally measurable quantities -- capable of clearly pinpointing where the Yukawa treatment breaks down -- has not been presented in the existing literature.

Monoclonal antibodies offer a particularly important case where these issues come to the forefront. From a biomedical perspective, they are essential therapeutics used for a wide variety of different treatments~\cite{paul2024cancer,lu2020development,carter2022designing}. From a more physical perspective, they can be seen as anisotropic colloids whose charge distribution is dictated by their amino acid sequence: some antibodies are relatively uniform in charge, while others display highly heterogeneous patterns that give rise to electrostatic patches and directional attractions~\cite{stradner2020potential,Roberts2014, Singh2014, Yadav2012,pineda2025patchy}. Despite this complexity, Yukawa-based models remain widely employed, as they provide a convenient way to approximate thermodynamic quantities such as second virial coefficients or to gain qualitative insights into phase behavior~\cite{Roberts2014, gulotta2024combining, dear2019x,chowdhury2023characterizing}. In such frameworks, the antibody structure is represented using a bead-based model derived from its amino acid sequence, in which each bead contributes to the electrostatic interactions through a point-Yukawa bead potential. However, there are already strong indications that such a model is not suitable for describing certain aspects of the system. 
%in partiuclar dynamical properties such as the zero shear viscosity. 
In fact, a recent study has shown that while the solution structure of relatively homogeneously charged antibodies can be captured reasonably well, the same models fail to reproduce experimental viscosity, unless explicit Coulomb interactions are included~\cite{camerin2025electrostatics}. This raises the broader question of how far the point-Yukawa bead description can be trusted for such systems~\cite{besley2023recent}.

It is therefore essential to systematically assess the validity of bead-based models dressed with point-Yukawa sources in the context of complex biomolecules. Here, we focus on two prototypical cases: one antibody with a relatively uniform charge distribution and another with pronounced heterogeneity. These situations not only occur naturally among different antibodies, but are also representative of a much wider class of biomolecules and colloidal particles where charge anisotropy plays a central role~\cite{gnidovec2025anisotropic,kim2024surface,guo2021quantifying,lund2016anisotropic,christians2024formalizing,weimar2024effective}.

Therefore, in this work, we develop a systematic, multiscale strategy to probe the reliability of point-Yukawa-based descriptions for antibodies and, more generally, for heterogeneously charged colloidal particles through direct comparison with experimentally measurable quantities that report on their solution structure and interactions.
Starting from atomistic simulations, we construct a coarse-grained representation of the antibodies, where each amino acid is replaced by a bead, and then use constant-pH Monte Carlo simulations with titration moves to determine the charge state of each amino acid and thus the charge distribution on the whole structure of the antibody.
These models are then used to perform many-body simulations under two alternative treatments of electrostatics: a bead model based on point-Yukawa potentials, and explicit Coulomb interactions including counterions and buffer ions. By applying this framework to the two mentioned cases, we directly test the limits of the point-Yukawa bead approach across systems of increasing complexity. 
We show how heterogeneous charge patterns disrupt the validity of the point-Yukawa bead approximation, effectively altering the single-molecule charge representation, and, consequently, the interactions and collective behavior. Crucially, by directly comparing simulation with experimental measurements -- such as static structure factors from x-ray scattering, and osmotic compressibility and  apparent hydrodynamic radius from static and dynamic light scattering -- we demonstrate that only the explicit charge approach reproduces the observed behavior, while the point-Yukawa bead description systematically fails when charge heterogeneity is pronounced. Theoretical analysis from liquid state theory corroborate this evidence. 
In doing so, this study goes beyond conventional point-Yukawa bead descriptions and introduces a strategy to capture how molecular-level charge heterogeneity shapes collective behavior, enabling predictive modeling of complex biomolecular and colloidal systems.

\begin{figure*}[t!]
\centering
\includegraphics[width=\textwidth]{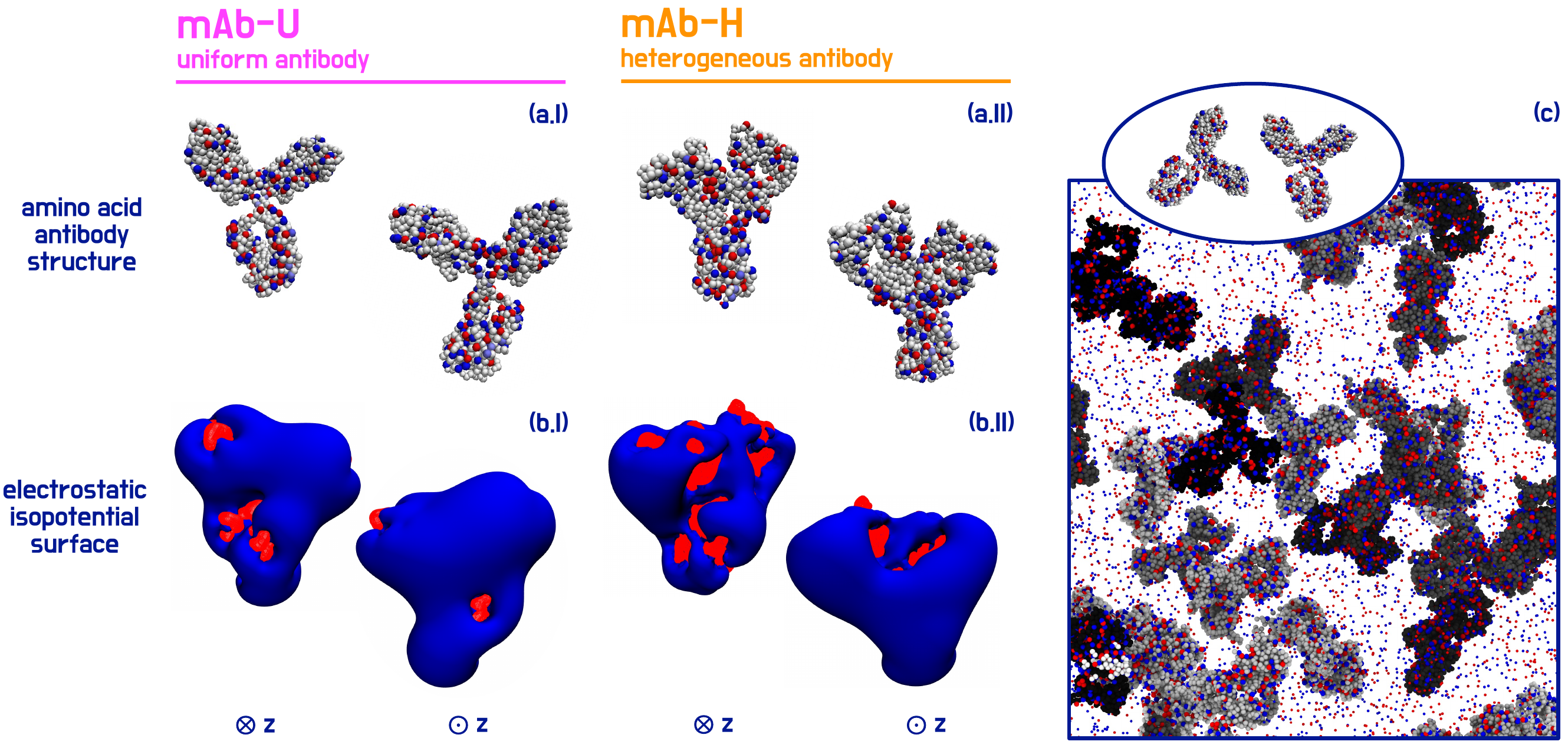}
\caption{\textbf{Antibodies under investigation.} (a) Amino acid representation of the antibody and (b) respective electrostatic isopotential surface (EIS) at $\pm 0.75 k_\mathrm{B}T/e$ calculated with a Poisson-Boltzmann solver for (I) mAb-U and (II) mAb-H from two different perspectives, indicated by $\otimes z$ and $\odot z$. (c) Representative simulation snapshot showing multiple antibodies at the amino acid level. Colored are the beads with charges $\pm 1$, while different shades of grey are for different molecules. For visual clarity, ions and counterions are not drawn to scale, and their number is reduced relative to the actual simulations. The inset shows two representative antibodies taken from the simulation box. In all panels, blue (red) represents positive (negative) charges.} 
\label{fig:antibodies}
\end{figure*}

\section{Results}

\subsection{Antibodies with different charge distributions}

We focus on two distinct monoclonal antibodies whose structure is depicted in Figure~\ref{fig:antibodies}a from two different viewpoints. Details of these macromolecules and their most important physical properties are given in Materials and Methods.
The reported snapshots show the coarse-grained antibodies, with each bead representing an amino acid colored according to its charge. Throughout the paper, we use blue for positive charges and red for negative ones. Such a coarse-grained representation is obtained by combining simulations with different degrees of coarse-graining techniques. In particular, atomistic simulations are used for obtaining an equilibrium structure, while constant pH Monte-Carlo simulations are used for assigning the charges to each amino acid, given the specific conditions of pH and ionic strength in the solution. More details on this protocol are provided in Methods. 

The charge distribution on each antibody is better visualized by drawing the corresponding electrostatic isopotential surface, which is obtained by applying a Poisson-Boltzmann solver to the previously discussed amino acid structures (see Methods). The surfaces at $\pm 0.75 k_\mathrm{B}T/e$ are shown in Figure~\ref{fig:antibodies}b, where $e$ is the elementary charge unit. It is precisely by this representation that the differences between antibodies become clear. In fact, despite both being characterized by an overall positive charge, the antibody at the left presents only a few negatively charged spots, which remain fully surrounded by positive charged domains. On the contrary, the one at the right presents a markedly inhomogeneous charge distribution with predominantly negative regions emerging in different parts of the protein.
Therefore, based on their charge distributions, we denote the two antibodies (mAb) as \textit{uniform} (mAb-U) and \textit{heterogeneous} (mAb-H).

\subsection{Experimental characterization and comparison}

The solution properties of both antibodies were characterized with static and dynamic light scattering (SLS and DLS), as well as with small-angle X-ray scattering (SAXS) measurements, and the results are summarized in Figures S2 and S3 in the Supplementary Information (SI) file. Here, we first focus on the comparison of the concentration dependence of the osmotic compressibility, expressed by the asymptotic low-$q$ value of the static structure factor $S(0) = M_\mathrm{w,app} /M_\mathrm{w}$, where $M_\mathrm{w,app}$ is the apparent weight-average molecular weight and $M_\mathrm{w}$ is the mass of the antibody, and of the apparent hydrodynamic radius $R_\mathrm{h,app}$. Figure~\ref{fig:expscharact} thus shows $S(0)$ (left panel) and  $R_\mathrm{h,app}$ (right panel) for the two antibodies at low concentrations $c$, i.e., in the second-virial regime. Under these conditions, we expect a linear concentration dependence expressed by
\begin{align}
	\label{eqn:kI}
	S(0) = 1 - k_\mathrm{I} c
\end{align}
and
\begin{align}
	\label{eqn:kD}
	R_\mathrm{h,app} = R_{\mathrm{h},0} (1 - k_\mathrm{D} c),
\end{align}
where $k_\mathrm{I}$ and $k_\mathrm{D}$ are the interaction coefficients describing the initial concentration dependence of the osmotic compressibility and the collective diffusion coefficient, respectively \cite{Corti1981}. 

\begin{figure*}[t!]
\centering
\includegraphics[width=\textwidth]{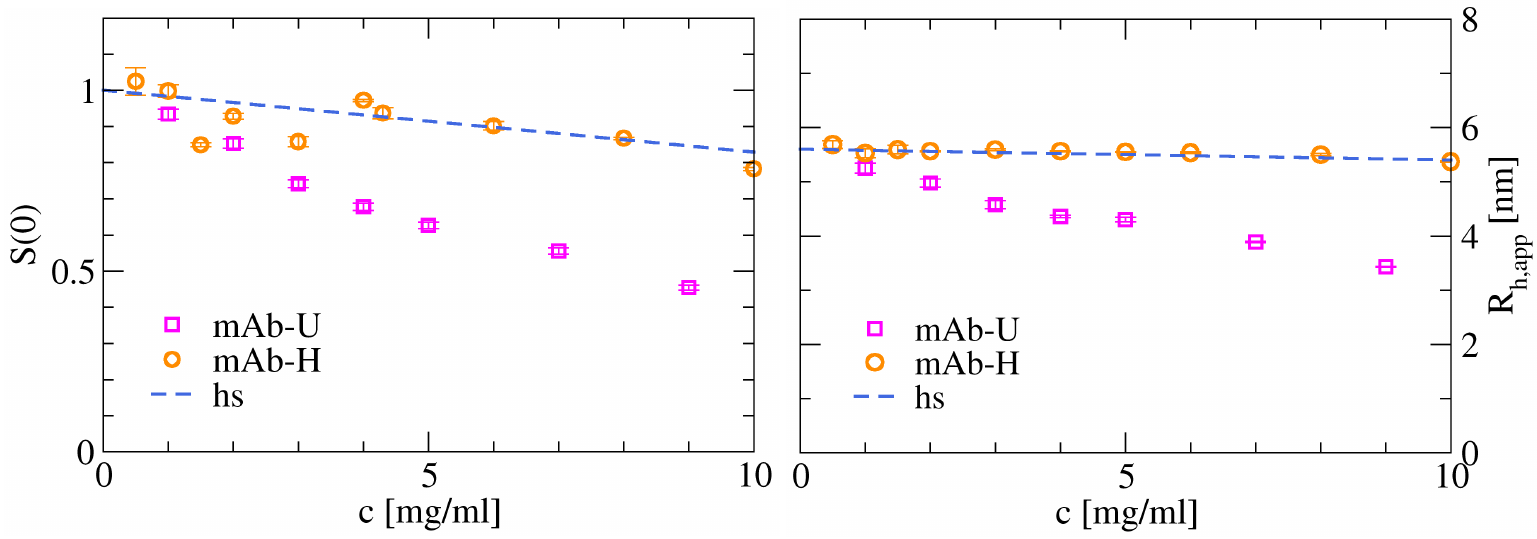}
\caption{\textbf{Experimental characterization.} (Left) Osmotic compressibility $S(0)$ and (right) apparent hydrodynamic radius $R_\mathrm{h,app}$ as a function of the antibody concentration $c$ for mAb-U (squares) and mAb-H (circles). Also reported is the theoretical prediction for excluded volume interactions (hs, dashed line).}
\label{fig:expscharact}
\end{figure*}

We observe significant differences between the two antibodies already at these low concentrations. In particular, mAb-U exhibits a behavior that is typical of highly charged particles at low ionic strength, where both $k_\mathrm{I}$ and $k_\mathrm{D}$ are quite large and $S(0)$ and $R_\mathrm{h,app}$ decrease significantly with increasing concentration. Moreover, under these conditions, the second-virial regime is quite narrow and requires measurements at very low concentrations to obtain accurate values for the interaction coefficients. Despite having a similar net charge (see Table \ref{tab:properties}), the behavior observed for mAb-H is quite different. The concentration dependence is much weaker, and all concentrations shown in Figure~\ref{fig:expscharact} lie within the dilute regime, indicating that a description truncated after the second virial coefficient provides an adequate approximation. In fact, the interaction coefficients are well described by a hard-sphere model (dashed line). For obtaining the corresponding hard-sphere relation, we use 
\begin{align}
	\label{eqn:kI-B2}
	k_\mathrm{I} = 2 \frac{N_\mathrm{A}}{M_\mathrm{w}}B_2
\end{align}
with $N_\mathrm{A}$ being Avogadro's number and
$B_2$ is the second virial coefficient given by
\begin{align}
	\label{eqn:B2}
	B_2 = 2 \pi  \int_{0}^{\infty}  \left[1 - e^{-\beta U(r)}\right] r^2 \,dr
\end{align}
where $\beta=1/k_\mathrm{B}T$ and $U(r)$ is the rotationally averaged potential of mean force~\cite{Corti1981, PeterLectures, Bratko2002}. For hard spheres, $B_2$ is given by $B_2^\mathrm{hs} = (2 \pi/3)\sigma_\mathrm{hs}^3$, where $\sigma_\mathrm{hs}$ is the hard sphere diameter. For $k_\mathrm{D}$ the corresponding hard sphere value is given by $k_\mathrm{D} = 1.45 (\pi \sigma_\mathrm{hs}^3/6) N_\mathrm{A}/M_\mathrm{w}$ \cite{Corti1981, PeterLectures}. For antibodies, we previously found that excluded volume interactions can be expressed by an effective hard sphere radius given by the radius of gyration of the antibody~\cite{gulotta2024combining}. Figure~\ref{fig:expscharact} shows that the concentration dependence is indeed well described by such a simple hard sphere model for mAb-H. 
This clearly indicates that at low concentrations and low ionic strength, where long-range charge interactions dominate, the soft repulsion originating from the net charge is significantly reduced by additional contributions related to the heterogeneous charge distribution with well-defined oppositely charged patches on mAb-H. 

Similar differences in the concentration dependence of the solution properties between the antibodies can also be observed in the SAXS results. These are shown for various concentrations in the SI and, for $c=20$ and $100$ mg/ml, as symbols in Figure~\ref{fig:sq_yuk}, with the left panel reporting results for mAb-U and the right one for mAb-H. The measured static structure factor $S(q)$ for mAb-U shows the typical pattern for particles with relatively high net charge at low ionic strength, with a strongly reduced forward scattering and a small but noticeable peak at $q=q^* \approx 0.04$ Å$^{-1}$ that corresponds roughly to the average nearest neighbor position given by the number density $n$, i.e. by $q^* \approx 2\pi/n^{-1/3}$~\cite{gulotta2024combining}. For higher concentrations, forward scattering becomes further reduced and the peak shifts to larger $q$-values. However, the measured $S(q)$ for mAb-H appears to be qualitatively different, with a less  reduced forward scattering and a small peak that is shifted to significantly larger $q$-values, almost overlapping with the weak peak seen at 100 mg/ml (see Figure~\ref{fig:sq_yuk}). This is once again more reminiscent of a behavior dominated by excluded volume interactions only, further supporting the observations made with SLS and DLS.

\subsection{Implicit versus explicit ions treatments}
\begin{figure*}[t!]
\centering
\includegraphics[width=\textwidth]{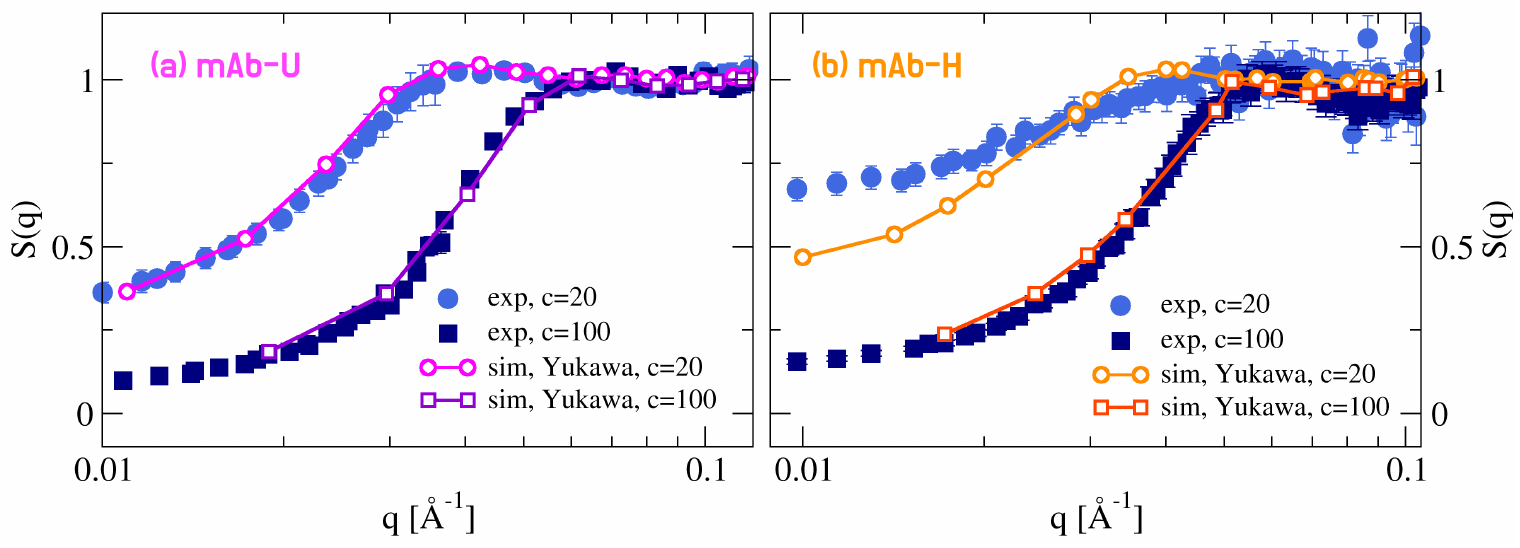}
\caption{\textbf{Solution behavior for the implicit ions model.} Static structure factors $S(q)$ for (a) mAb-U and (b) mAb-H for simulations (sim) with implicit ions models (Yukawa) and for SAXS experiments (exp) for $c=20$ and $100$ mg/ml.
}
\label{fig:sq_yuk}
\end{figure*}

By representing antibodies at the amino acid level, we can probe their many-body behavior while retaining a direct correspondence with their atomistic structure. This “weak” coarse-graining provides a model that is as close as possible to a real antibody, while minimizing artifacts associated with lower-resolution models and maintaining computational feasibility, which would be unattainable at the fully atomistic level.
An additional advantage is that these simulations can be quantitatively compared to experiments after an appropriate mapping of units. 
Former investigations on mAb-U by some of us suggest that this level of coarse graining may already be sufficiently detailed to capture experimentally measured static structure factors over a broad concentration range~\cite{polimeni2024multi}. In that study, the amino acid bead model was used as a basis to design and benchmark a coarse-graining strategy for concentrated antibody solutions. The simulations employed an implicit description of solvent and ions: beads interact via an effective pair potential of a point-Yukawa (screened Coulomb) form, supplemented by a short-range attraction that accounts for van der Waals and hydrophobic interactions (Eq.~\ref{eq:AA-Yuk}).
As shown in Fig.~\ref{fig:sq_yuk}a, with this approach, the simulated $S(q)$ (lines and symbols) closely match the experimental data at both low and high concentrations. 

However, when we adopt a similar approach for mAb-H (see Methods) and compute the corresponding static structure factor, we observe clear and systematic deviations from experiments (Fig.~\ref{fig:sq_yuk}b). The disagreement is most pronounced at the lowest concentration ($c = 20$ mg/ml), where electrostatic effects dominate, whereas at $c = 100$ mg/ml, where excluded-volume and short-range attractions become more important, the agreement improves substantially.

Notably, mAb-U and mAb-H have very similar molecular weight, size, shape and net charge (Table~\ref{tab:properties}), yet they strongly differ in the way this charge is spatially distributed, as visualized by their electrostatic isopotential surfaces (Fig.~\ref{fig:antibodies}). Because our amino acid resolution is sufficiently fine to reflect this heterogeneity, these results suggest that the failure is not due to the degree of coarse-graining we employ, but rather to the interaction model itself. In fact, the implicit-ion, point-Yukawa bead description seems unable to capture the correct many-body behavior when the charge distribution becomes more patchy-like, as in mAb-H. By contrast, the nearly uniform charge distribution of mAb-U poses no such challenge.

\begin{figure}[hb!]
\centering
\includegraphics[width=0.48\textwidth]{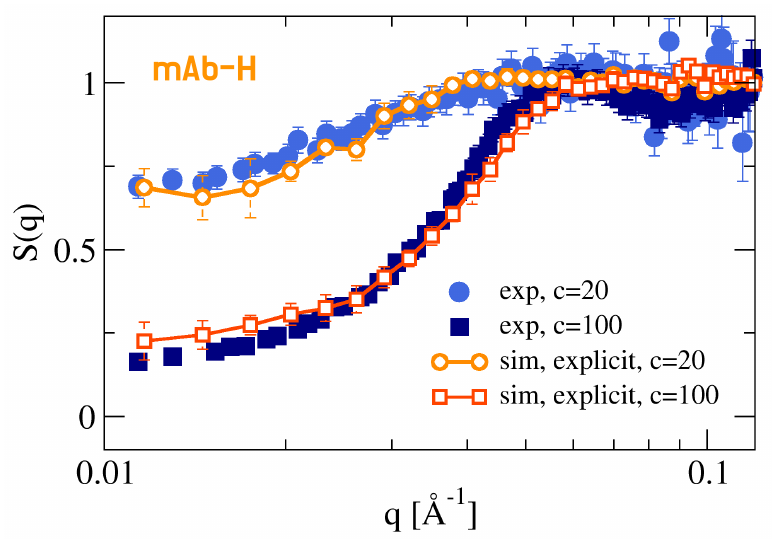}
\caption{\textbf{Solution behavior for the explicit ions model.} Static structure factors $S(q)$ for mAb-H for simulations (sim) with explicit ions models (Coulomb) and for SAXS experiments (exp) for $c=20$ and $100$ mg/ml. 
%\ez{with and without buffer ions. do we have it?  do we merge Fig.1 and 2?}
The experimental data points are the same as in Fig. \ref{fig:sq_yuk}b.}
\label{fig:sq_coul}
\end{figure}

To test this hypothesis directly, we performed amino acid level simulations of mAb-H with explicit charges and ions. Charged amino acids interact via the bare Coulomb potential (Eq.~\ref{eq:AA-Coul}), and ions and counterions are added to mimic the presence of the buffer and ensure charge neutrality (see Methods). A snapshot of this simulation setup is reported in Figure~\ref{fig:antibodies}c. Because of the long-range nature of Coulomb interactions and the large number of particles, these simulations are extremely demanding and suffer from non-negligible noise in the low-$q$ regime. To obtain sufficiently converged statistics for the compressibility, we therefore ran at least three independent replicas at both concentrations. The resulting static structure factors obtained are reported in Figure~\ref{fig:sq_coul}. They show excellent agreement with experiments at both concentrations, demonstrating a marked qualitative improvement over the point-Yukawa bead description. This confirms that an accurate treatment of electrostatics is essential to recover the correct collective behavior when the molecule surface charge distribution is highly heterogeneous.

\subsection{Implications at the single-molecule level}

The previous results clearly demonstrated the important role of charge heterogeneities in controlling the structure of antibody solutions. We now attempt to obtain a better understanding of the above findings. The aim is to provide an essential and clear picture of the differences that arise by treating the molecules with an implicit or explicit ions modeling, and to clarify why the implicit treatment remains adequate for antibodies with a nearly homogeneous charge distribution, yet fails when the charge becomes strongly patchy.

An interesting perspective on the reasons why our implicit ions model yields an underestimation of the compressibility for mAb-H, especially at low concentrations, can be gained by revisiting the electrostatic isopotential surfaces. In fact, those initially reported in Figure~\ref{fig:antibodies}b were obtained by numerically solving the Poisson-Boltzmann equation, which explicitly accounts for the spatial distribution of counterions and the nonlinear screening effects near charged domains, thereby providing a relatively accurate description of the charge distribution. It is thus interesting to compare these results with those obtained from a calculation based on an implicit ions model using a bead model dressed with point Yukawas for the same charge distribution on the antibody.

\begin{figure*}
\centering
\includegraphics[width=\textwidth]{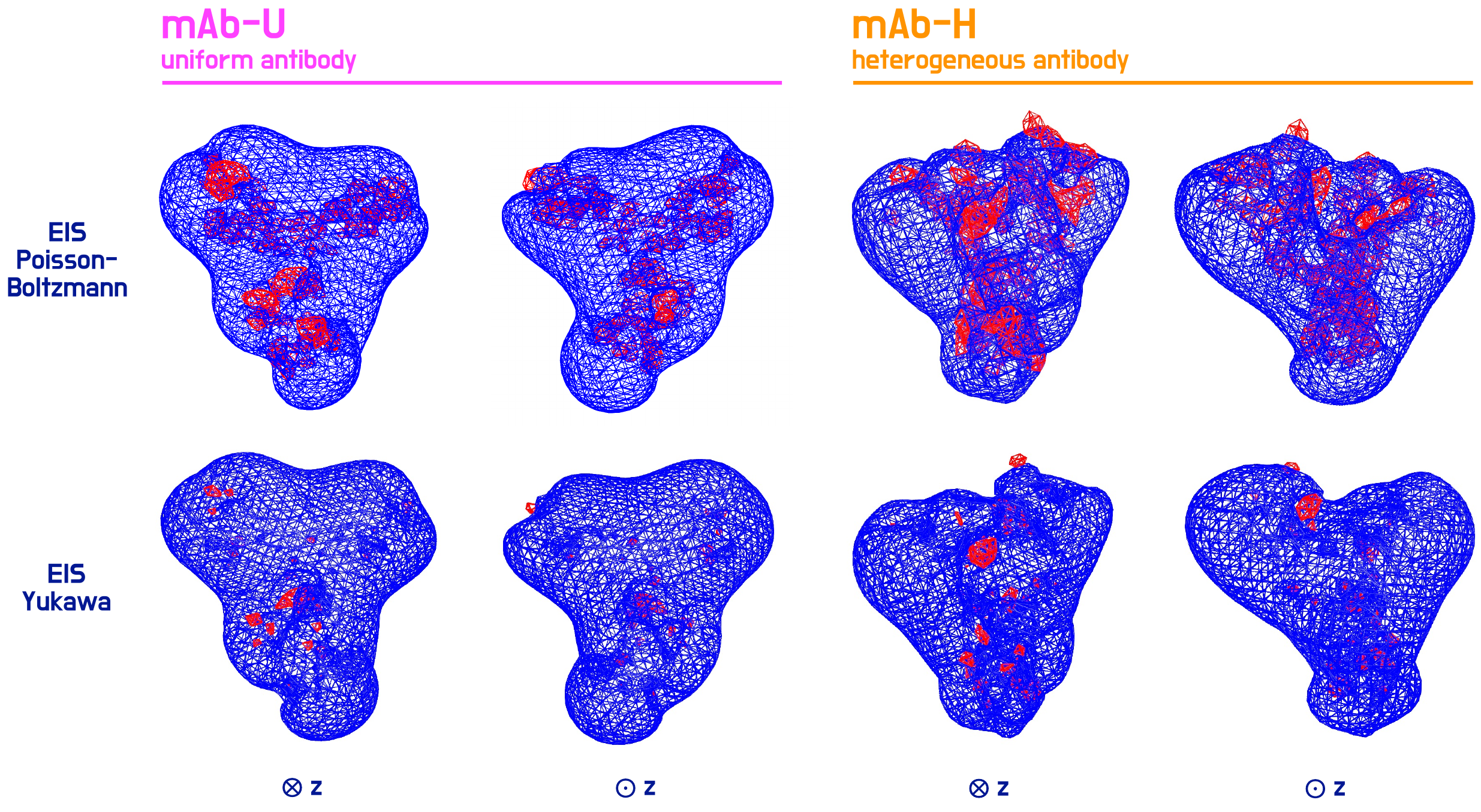}
\caption{\textbf{Comparison between PB and Yukawa representation.} Mesh representation of the electrostatic isopotential surfaces (EIS) at $\pm 0.75 k_\mathrm{B}T/e$ accounting for (top) Poisson-Boltzmann and (bottom) Yukawa representations for (left) mAb-U and (right) mAb-H from two different perspectives, indicated by $\otimes z$ and $\odot z$.
}
\label{fig:cfr_esi}
\end{figure*}

%To assess how far our model departs from the actual charge distribution, 
We therefore generate an analogous representation in which the Yukawa potential replaces the Poisson–Boltzmann treatment at each point of the isopotential surface (see Methods). The results are shown in Figure~\ref{fig:cfr_esi} for mAb-U and mAb-H, respectively, compared to the surfaces previously obtained. In this case, we use a mesh representation to more clearly highlight the differences between the two methods.
The analysis of the newly obtained isopotential surfaces reveals that the negative potential domains, although still present, are significantly reduced compared to the more accurate Poisson-Boltzmann treatment. This effect is particularly pronounced for the more heterogeneously-charged antibody, where the negative regions appear even more fragmented and smaller. Overall, the electrostatic isopotential surface thus appears much more homogeneous and dominated by the overall positive net charge for the point-Yukawa calculation, indicating that antibody–antibody attraction caused by oppositely charged patches is significantly reduced under these conditions. This subsequently leads to a stronger reduction of the compressibility at low concentrations.

In order to rationalize these differences, it is instructive to examine more closely the fundamental equations underlying the two representations. For a single antibody described at the amino acid level, the  electrostatic potential $\psi({\bf r})=k_\mathrm{B}T\phi({\bf r})/e$ satisfies the Poisson–Boltzmann equation in the water solvent region $V_\mathrm{w}$ where $\phi({\bf r})$ is determined by
\begin{equation}
\nabla^2 \phi(\mathbf{r}) = \ell_\mathrm{D}^{-2} \sinh[\phi(\mathbf{r})], \quad \mathbf{r}\in V_\mathrm{w},
\end{equation}
with $\ell_\mathrm{D}$ the Debye screening length.
Within the antibody domain $V_\mathrm{mAb}$ only the fixed charges located at the amino acid positions contribute:
\begin{equation}
\frac{\varepsilon_\mathrm{mAb}}{\varepsilon_\mathrm{w}} \nabla^2 \phi(\mathbf{r}) = -4\pi \ell_\mathrm{B} \sum_{i=1}^{N_\mathrm{q}} q_i  \delta(\mathbf{r}-\mathbf{r}_i), \quad \mathbf{r}\in V_\mathrm{mAb},
\end{equation}
where $\varepsilon_\mathrm{mAb}=1$ and $\varepsilon_\mathrm{w}=78.7$ are the relative permittivities within the antibody interior and of the solvent, respectively. Furthermore, $\ell_\mathrm{B}=\beta e^2/(4\pi\varepsilon_0\varepsilon_\mathrm{w})$ is the Bjerrum length of the solvent, $N_\mathrm{q}$ is the number of amino acids, and $q_i$ is the total charge of the $i$-th bead in the amino-acid representation of the antibody.
At the interface between these two regions, appropriate boundary conditions are imposed to ensure the continuity of both the electrostatic potential and the normal component of the displacement field. These equations inherently account for the exclusion of mobile ions from the low-dielectric interior of the antibody and the resulting image-charge effects that arise at the dielectric boundary~\cite{hunter1987foundations}.
%ADD REFS \cite{xxx}.

By contrast, the point-Yukawa bead description can be derived as a simplified limit of the PB treatment under the following assumptions:
(i) the dielectric contrast between the antibody and the solvent is neglected,
(ii) ions are allowed to freely penetrate the molecular interior, and
(iii) the electrostatic potential remains small in magnitude, i.e., $|\phi({\bf r})| \ll 1$.
Under these approximations, the potential satisfies the linearized Poisson–Boltzmann equation,
\begin{gather}
\nabla^2\phi({\bf r}) = \ell_\mathrm{D}^{-2}\phi({\bf r}) - 4\pi\ell_\mathrm{B}\sum_{i=1}^{N_\mathrm{q}} q_i  \delta({\bf r}-{\bf r}_i), \quad {\bf r}\in V_\mathrm{w}\cup V_\mathrm{mAb}, \label{eq:LPB}
\end{gather}
whose solution reads
\begin{equation}
\phi({\bf r}) = \ell_\mathrm{B}\sum_{i=1}^{N_\mathrm{q}} q_i  \frac{e^{-|{\bf r}-{\bf r}_i|/\ell_\mathrm{D}}}{|{\bf r}-{\bf r}_i|}. \label{eq:ptYukpot}
\end{equation}
This expression, while analytically tractable, does not solve the full nonlinear Poisson–Boltzmann equation. It instead assumes that all charges interact through a homogeneous dielectric medium uniformly populated by screening ions, so that the total potential can be expressed as a linear superposition of Yukawa terms. This is the representation underlying the isopotential surfaces shown in the lower panels of Figure~\ref{fig:cfr_esi}.

At this point, the specific charge distribution of the antibody becomes crucial, as the simplifications inherent in the Yukawa picture become increasingly inadequate with increasing electrostatic heterogeneity. In the case of mAb-H, the resulting electrostatic fields display pronounced spatial variations near the molecular surface, where dielectric mismatch and ion exclusion effects are strongest. Moreover, as demonstrated in simpler charged systems~\cite{gnidovec2025anisotropic},
%add refs Jeffrey
spatial inhomogeneities enhance the relevance of higher-order multipole contributions to the potential. These multipole moments require further corrections in a screened environment, such as an aqueous solution, thereby amplifying deviations from the idealized point-Yukawa form. Collectively, these considerations exemplify why a simple point-Yukawa bead representation fails to capture the complex electrostatic landscape of heterogeneously charged antibodies. More generally, deriving effective electrostatic interactions for molecules with unconventional geometry and strongly anisotropic charge patterns remains a largely open theoretical problem, and will require substantial future effort to properly incorporate higher-order multipole contributions and screening non-linearities, although there are some recent developments in this direction~\cite{everts2020,gnidovec2025anisotropic,obolensky2021rigorous,ter2025machine}.

\subsection{Insights from the potential of mean force between antibodies}

We finally turn to analyze the effective orientationally-averaged pair potential, or potential of mean force (PMF) $U(r)$, between antibodies, which is another descriptor where differences between treatments becomes evident. 
Moreover, using $U(r)$, we can also calculate the corresponding interaction coefficients $k_\mathrm{D}$ and $k_\mathrm{I}$ and directly test these values against the experimental data obtained at low concentrations.
While for mAb-U, the PMF based on the amino acid model with Yukawa interactions was shown to be successful in solution properties such as the osmotic compressibility and the apparent hydrodynamic radius~\cite{gulotta2024combining}, we expect that significant deviations will occur for mAb-H.

\begin{figure}[hb!]
\centering
\includegraphics[width=0.48\textwidth]{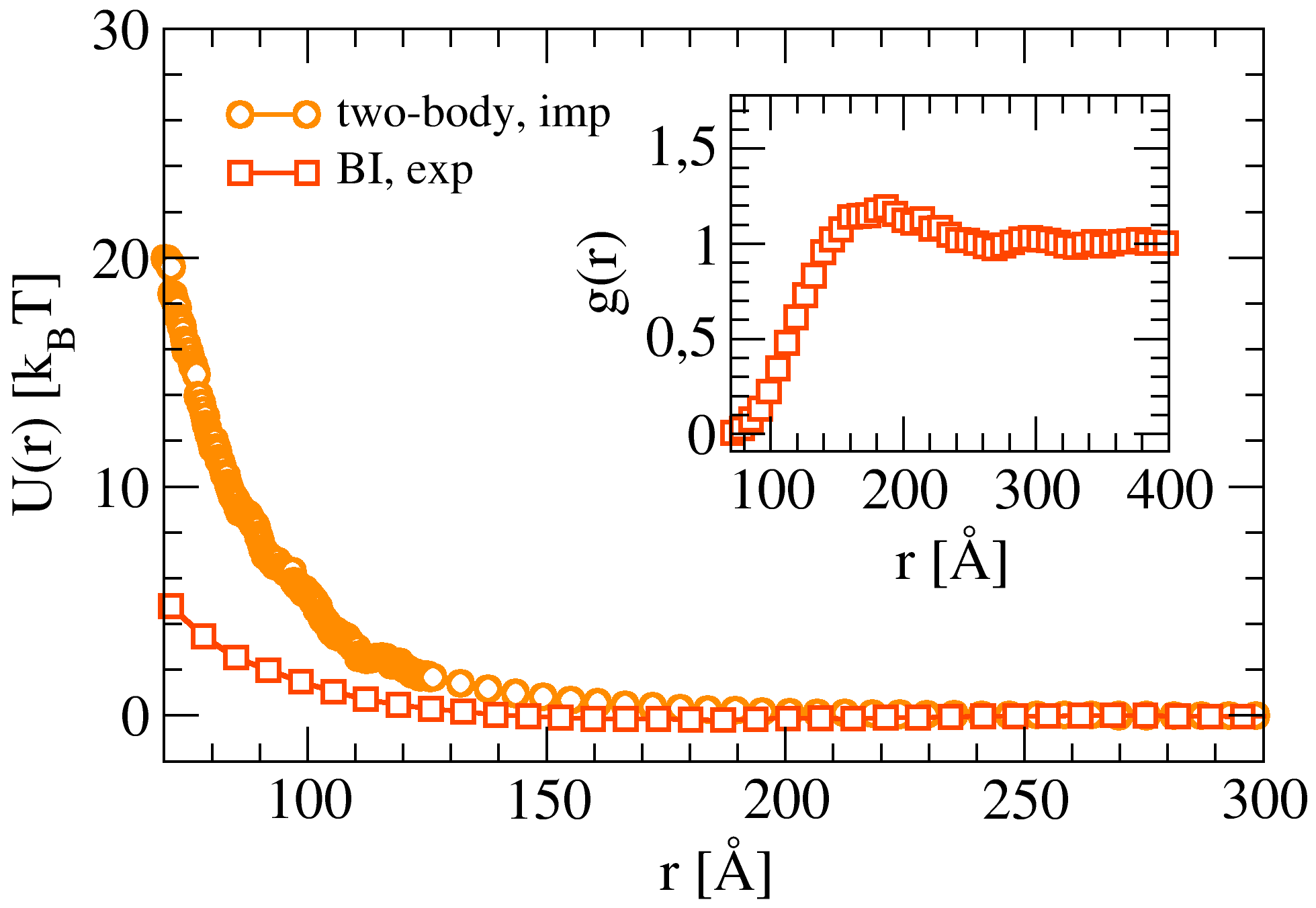}
\caption{\textbf{Inter-protein interactions for the heterogeneous antibody.} Effective interaction potential $U(r)$ between two mAb-H antibodies calculated by using the Yukawa potential (circles, imp), and obtained by Boltzmann inversion (BI) (squares, exp) from the radial distribution function $g(r)$ (shown in the inset) of simulations at low concentration where explicit Coulomb interactions were used. 
}
\label{fig:effpot}
\end{figure}

\begin{figure*}[th!]
\centering
\includegraphics[width=\textwidth]{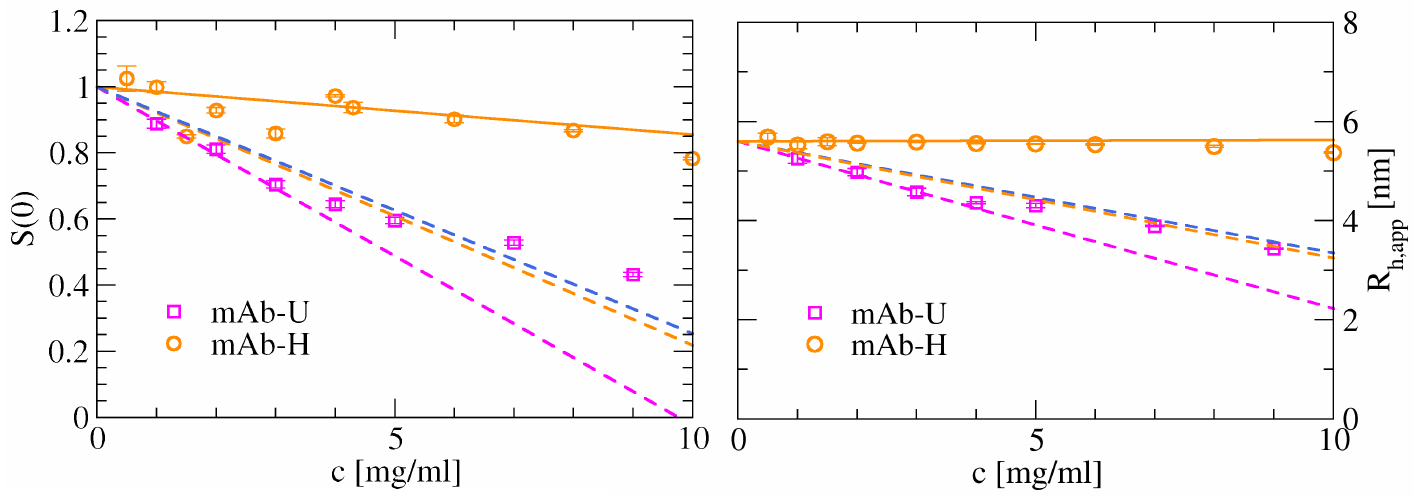}
\caption{\textbf{Test of potential of mean force.} (Left) Osmotic compressibility $S(0)$ and (right) apparent hydrodynamics radius $R_\mathrm{h,app}$ as a function of the antibody concentration $c$ for mAb-H and mAb-U. Also reported are the theoretical predictions using the potential of mean force for implicit (magenta dashed line: mAb-U; orange dashed line: mAb-H; blue dashed line: homogeneously charged antibody with net charge $Z = +24$) and explicit (orange solid line: mAb-H) ions models.
}
\label{fig:kI-kD}
\end{figure*}

We thus calculate the PMF for the more heterogeneous antibody for implicit and explicit ions using two different approaches. For the implicit ions modeling, we use a combination of the Widom insertion method and Umbrella Sampling (see Methods). In the presence of explicit charges, calculating effective potentials becomes particularly challenging. This difficulty arises not only from the long-range nature of electrostatic interactions but also from the need to properly control charged species in a dilute environment. Although innovative computational methods are being developed to address this issue~\cite{krucker2025potential}, their implementation lies beyond the scope of the present work.
For this reason, we make use of the Boltzmann inversion (BI) method, which relies on the ``inversion" of the radial distribution function $g(r) \approx$ exp$[-\beta U(r)]$ for a dilute suspension of objects. To this aim, we run simulations of several antibodies in solution at different concentrations $c=10-20$ mg/ml, which are the lowest concentrations that still ensure the feasibility of a simulation with a large ensemble of charged antibodies and small ions. To reduce statistical noise, we run two 
%\ps{were there not more replicas finally?} 
replicas with different initial configurations. The results for simulations with implicit and explicit ions for mAb-H are shown in Figure~\ref{fig:effpot}, with the inset displaying the radial distribution function used to calculate the effective potential via the BI method. Consistent with the above findings, the implicit ions model predicts markedly stronger repulsion than the explicit ions treatment. In the latter case, the interactions remain repulsive, but the strength of the repulsion is substantially reduced at short distances.

From the obtained PMF, we can then calculate $k_\mathrm{I}$, using Equations~\ref{eqn:kI-B2} and \ref{eqn:B2}, while $k_D$ is obtained from liquid state theory, as described in Methods. Figure~\ref{fig:kI-kD} summarizes our findings by reporting the concentration dependence of $S(0)$ and of $R_\mathrm{h, app}$ for different models (lines) and by comparing them with the experimentally measured data (squares for mAb-U and circles for mAb-H). As previously reported~\cite{gulotta2024combining}, an implicit ions treatment with a Yukawa potential for the charge contributions (magenta dashed lines) reproduces the experimental results obtained for mAb-U quite well. However, Figure~\ref{fig:kI-kD} also shows that the second-virial regime for such a highly charged antibody is quite limited to $c \lesssim 4$ mg/ml, which makes a precise measurement of the interaction coefficients difficult and can result in systematic errors when measuring at too high concentrations.

For mAb-H, the PMF obtained for the implicit ions model is not able to reproduce the experimental data (orange dashed line), supporting the findings made when trying to calculate the full structure factor measured with SAXS (see Fig. \ref{fig:sq_yuk}). It is, in fact, quite interesting to compare the results for the implicit ions model with the predictions for a homogeneously charged antibody with a net charge of +24, which are also shown in Figure~\ref{fig:kI-kD} (blue dashed line). The predictions for $k_\mathrm{I}$ and $k_\mathrm{D}$ both almost completely coincide with those for the heterogeneous charge model when using a Yukawa potential. This finding is in line with the observations made when looking at the isopotential surfaces in Figure~\ref{fig:cfr_esi}, where the negative patches almost disappear and where the isopotential surface for 0.75 $k_\mathrm{B}T/e$ appears almost homogeneous. The long range part of the PMF that dominates the volume integrals needed to calculate $k_\mathrm{I}$ and $k_\mathrm{D}$ are thus mainly determined by the electrostatic repulsion related to the net charge, and the additional attraction caused by oppositely charged patches diminishes when using an implicit ions model. In contrast, the PMF obtained for the primitive model with explicit ions (orange solid line) is capable to reproduce the initial concentration dependence of $S(0)$ and $R_\mathrm{h,app}$ almost quantitatively.

\begin{figure*}[th!]
\centering
\includegraphics[width=\textwidth]{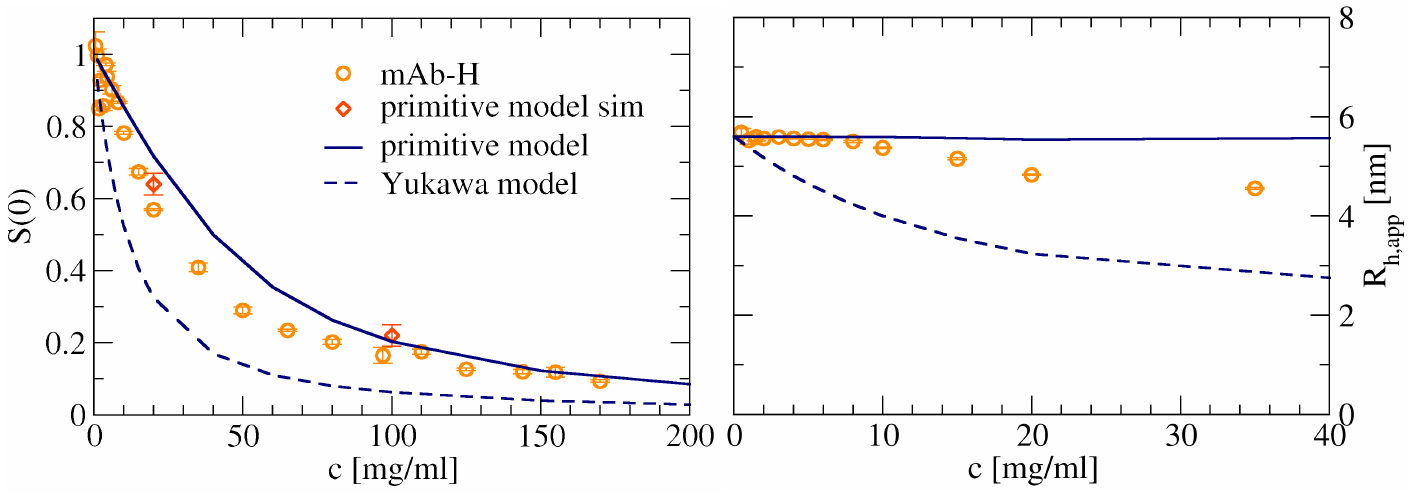}
\caption{\textbf{Concentration dependence of solution properties.} (Left) Osmotic compressibility $S(0)$ and (right) apparent hydrodynamics radius $R_\mathrm{h,app}$ as a function of concentration $c$ for mAb-H. Also reported are the theoretical predictions using the potential of mean force for implicit (dark blue dashed line) and explicit (blue solid line) ions and the simulation results for $S(0)$ from the explicit ions simulations (diamond symbols).}
\label{fig:S0-Rh}
\end{figure*}

It is only at intermediate concentrations (10 mg/ml $\lesssim c \lesssim$ 100 mg/ml) that the model deviates. This is further illustrated in Figure~\ref{fig:S0-Rh}, 
%Here we calculate $S(0)$ and $R_{h,app}$ as described in the methods section. 
where it appears that the PMF obtained from the primitive model simulations through BI is too attractive to correctly reproduce the experimental data at high concentrations. There are two main possible reasons for such behavior. First, in the use of BI to obtain the PMF, we assume that we can apply a dilute gas approximation, where the pair distribution function is given by $g(r) \approx$ exp$[-\beta U(r)]$. However, using an explicit ions model with a weakly coarse-grained antibody, we are limited in extending our simulations to very low concentrations, and this assumption may thus no longer hold. Given the quite good agreement between calculated and measured properties at low concentrations (Fig. \ref{fig:kI-kD}), where in particular $k_\mathrm{D}$ is very sensitive to small variations in the PMF, we are fairly confident that the chosen conditions do not result in larger errors in our calculation of $U(r)$. 
Another possible reason for the failure relates to the assumption of a PMF that is independent of concentration.
%Another possible reason for the failure of HMSA to reproduce the measured data at intermediate concentrations could be the assumption of a PMF that is independent of concentration. 
For low ionic strength buffers, as used here, the overall ionic strength of the solution will also depend on the concentration, and we thus expect variations in the strength of the electrostatic contributions to the PMF with increasing antibody concentrations. While these effects would be less important at very high concentrations, where excluded volume effects dominate, their influence would be largest at intermediate concentrations. The good agreement of the simulation results with the measured data at 20 mg/ml and 100 mg/ml, also shown in Figure~\ref{fig:S0-Rh}, indicates the chosen coarse-grained explicit ions model is indeed capable of reproducing the solution structure accurately. This suggests that the use of a constant PMF obtained at low concentrations is the main source of error in the calculation of $S(0)$ and $R_\mathrm{h,app}$. As we have no analytical model that would allow us to include an explicit ionic strength dependence of the PMF in our calculations of $S(q)$ and $H(q)$, the only possible approach would be an iterative inversion of the simulated $g(r)$ using an
%HMSA or another 
appropriate closure relation at all concentrations investigated to obtain an effective concentration-dependent PMF~\cite{Law2010}.

%\subsection{Comparison with experiments $k_D$}

\section{Discussion}\label{sec12}

Our combined multidimensional analysis demonstrates that the validity of screened-Coulomb models for antibodies is highly dependent on the spatial distribution of charges. For the antibody with a relatively uniform charge distribution (mAb-U), implicit ions simulations using a point-Yukawa bead representation reproduce scattering and thermodynamic data across the explored concentration range. This indicates that, when charge heterogeneity is weak, a linearized mean-field treatment -- where ion-exclusion and image-charge effects are neglected -- provides an accurate representation of the effective interactions.

In contrast, for the antibody with marked charge heterogeneity (mAb-H), the same approach fails systematically. The experimental compressibility measured by SLS, the concentration dependence of the apparent hydrodynamic radius extracted from DLS, and the low-q behaviour of SAXS all show discrepancies relative to the implicit ions predictions. Explicit ions simulations at the amino-acid level, although computationally more demanding, recover the experimental static structure factors across concentrations and yield a potential of mean force that produces a behavior in line with that measured experimentally.
The comparison of isopotential surfaces between the Poisson–Boltzmann and the point-Yukawa bead model also illustrates the origin of these deviations. In mAb-H, Poisson–Boltzmann preserves the spatially extended negative surface domains, whereas the point-Yukawa bead approximation attenuates them and effectively homogenizes the electrostatic isopotential surface. This demonstrates that charge heterogeneity generates non-uniform electrostatic fields and non-linear screening responses that cannot be captured by the point-Yukawa model.

These findings also indicate that further research is needed when attempting to understand and predict thermodynamic properties, solution structure, diffusion and flow behavior of concentrated antibody solutions. While our amino acid level explicit ions simulations were able to reproduce the static structure factor and the osmotic compressibility also for the antibody with the heterogeneous charge distribution (mAb-H), such an approach is not feasible for other quantities such as the collective or self-diffusion coefficient or the viscosity. Therefore, it will be needed to develop more advanced coarse-graining strategies that retain the essential anisotropic features of the shape and charge distribution, but allow to extract dynamic quantities and effective potentials over a wide range of concentrations. Similarly, additional theoretical work will be needed for understanding the role of multipoles' contributions to the interaction potential of colloidal-scale particles with complex geometries.

In summary, our results show for a relevant study case that screened-Coulomb, implicit ions descriptions based on models dressed with point-Yukawa potentials are predictive only when the molecular surface charge is sufficiently uniform. Because these conclusions emerge consistently across scattering experiments, explicit-ion simulations and potential of mean-force analysis, they provide a solid criterion for selecting the appropriate electrostatic treatment in future studies.
More broadly, this implies that any attempt to rationalize solution behavior for complex biomolecules that display anisotropic surface charge -- such as antibodies, complex proteins and enzymes, or patchy colloidal and heterogeneously charged particles -- must explicitly account for how long-range electrostatics couples to local structural motifs. The integrated use of experiments and theoretical approaches presented here provides a route to establish this connection in a quantitative and transferable manner. Furthermore, the present study does not only applies to investigations of the solution structure but has far-reaching implication on its dynamical and rheological properties, where the balance between local patterning and electrostatics also plays a crucial role. These insights are also particularly valuable for practical applications such as the design of high-concentration therapeutic formulations, where accurate predictions of viscosity, intermolecular interactions, and solution stability are essential.
Ultimately, our results provide a defined framework for the analysis of complex colloids and biomolecules, supporting future investigations in crowded environments and establishing a more reliable basis for predictive modelling and rational exploitation of their functional properties.

\section{Materials and Methods}\label{sec11}
\small

\subsection{Antibodies} We investigate the solution behavior of two different antibodies. The first antibody, labeled mAb-U in this study, is an immunoglobulin G of subclass 1 (IgG1) named Actemra or Roactemra. The amino acid sequence of the antibody was obtained from the patent N. US20120301460. The atomistic structure of the antibody was built by homology modeling using the Molecular Operating Environment (MOE) software (Chemical Computing Group, Inc.). A more complete description of the homology model process can be found in Refs.~\cite{gulotta2024combining,polimeni2024multi}. A systematic experimental characterization of the main solution properties as a function of concentration and ionic strength has previously been reported in Ref. \cite{gulotta2024combining}, and the experimental data for the osmotic compressibility, apparent hydrodynamic radius and the static structure factors used in the current work were taken from there. 

The second antibody, Cetuximab, or Erbitux, is a recombinant chimeric human/mouse IgG1 monoclonal antibody that binds to the epidermal growth factor receptor. In this study, it is labelled as mAb-H.  The amino acid sequence for both the light and heavy chains of Cetuximab are reported in the IMGT database (www.imgt.org) and the drug bank (www.drugbank.ca), and its primary sequence has also been reported by Dubois et al. \cite{Dubois2008}. The crystal structure of the antigen binding fragment (Fab) has been reported by Li et al. \cite{Li2005} and is referenced in the RCSB Protein Data Bank. The atomistic structure of the antibody was built by homology modeling using the Molecular Operating Environment (MOE) software (Chemical Computing Group, Inc.).
%The \textsc{pdb} structure of both antibodies is provided in the indicated repository. 
Their most important physical properties are summarized in Table \ref{tab:properties}.

\begin{table}[h!]
\caption{\footnotesize Physical properties of mAb-U and mAb-H. Shown are the molecular weight $M_\mathrm{w}$, the hydrodynamic radius $R_\mathrm{h}$ extrapolated to infinite dilution measured by DLS, the radius of gyration $R_\mathrm{g}$ measured by SAXS, the isoelectric point (IEP) measured by capillary isoelectric focusing (cIEF), and the net charge $Z_\mathrm{net}$ calculated using constant pH MC simulations as described below. Values are for the chosen ionic strength $I$ of 7 mM and pH 6.0.}
\centering
\begin{tabular}{| c | c | c | c | c | c |} 
 \hline
   & $M_\mathrm{w}$ [kDa] & $R_\mathrm{h}$ [nm] & $R_\mathrm{g}$ [nm] & IEP  & $Z_\mathrm{net}$  \\ [0.5ex] 
 \hline
 mAb-U & 148 & 5.4 & 5.0 & 9.18 & +31 \\ 
  \hline
 mAb-H & 152 & 5.6 & 5.3 & 8.10 & +24 \\ %[1ex] 
 \hline
\end{tabular}
\label{tab:properties}
\end{table}

\subsection{Sample preparation}

The samples used in this study were purchased commercially. Prior to experimentation, surfactant (polysorbate 80) was removed from the formulation using DetergentOUT Tween spin columns (G-Biosciences). Samples then underwent dialysis in 10,000 MWCO Slide-A-Lyzer cassettes (Thermo Fisher Scientific) to exchange into a basis buffer of 10 mM L-histidine at pH 6.0. Following buffer exchange, samples were concentrated to approximately 200 mg/mL using centrifugal concentrators (MilliPoreSigma). Samples were then filtered using 0.22 $\mu$m spin columns (Corning) and stored at $-80^{\circ}$C prior to analysis.

Dilution series and subsequent measurements were made using the same low ionic strength buffer (10 mM L-histidine at pH = 6.0). The buffer corresponding to the buffer of the stock solution was prepared by dissolving 5 mM of L-Histidine and 5 mM of Histidine-HCl Monohydrate (both Sigma-Aldrich, SE). The final pH of the buffer was adjusted to $6 \pm 0.05$ by the addition of a few microliters of hydrochloric acid (HCl, 0.1 M). This results in an ionic strength of 7 mM at the chosen pH = 6.

The samples at different concentrations were prepared by diluting the stock solution originally obtained. The frozen stock solution was thawed at room temperature ($\approx$ 20$^{\circ}$C) for $\approx$ 30 minutes, and then gently homogenized by using a micropipette. Once prepared, the samples were used for measurements, otherwise stored in a freezer at -80$^{\circ}$C. Before measurement, the concentration was measured via UV absorption spectroscopy, using a wavelength of $\lambda = 280$ nm and a specific absorption coefficient $\mathrm{E_{mAb, 1 \: cm}^{0.1\%, 280 \: nm}} =$ 1.51 $\mathrm{ml \cdot mg^{-1} \cdot cm^{-1}}$. 

\subsection{Experimental procedures} 

\textbf{(i) Small angle x-ray scattering.} Small Angle X-Ray Scattering (SAXS) measurements were performed with a pinhole camera system (Ganesha 300 XL, SAXSLAB) equipped with a high brilliance microfocus sealed tube and thermostated capillary stage. The accessible \emph{q}-range for these measurements was  from $5 \times 10^{-2} \lesssim q \lesssim 10$ nm$^{-1}$. Experiments were carried out at T = 25 $^{\circ}$C. All measurements were corrected for the background radiation, buffer in the capillary, mAb concentration, and transmission, resulting in a normalized scattering intensity $\bigg(\frac{d\sigma}{d\Omega}(q)\;c^{-1}\bigg)$. The  experimental structure factors ($S(q)$), were calculated using

\begin{equation}
  S(q) = \frac{\Bigg[\frac{d\sigma}{d\Omega}(q)\cdot c^{-1}\Bigg]}{\Bigg[\frac{d\sigma}{d\Omega}(q)\cdot c_{0}^{-1}\Bigg]_{\mathrm{FF}}}
\label{eq:9}
\end{equation}

\noindent where $\bigg[\frac{d\sigma}{d\Omega}(q)\cdot c^{-1}\bigg]$ is the normalized scattered intensity at higher protein concentration $c$ and $\bigg[\frac{d\sigma}{d\Omega}(q)\cdot c_{0}^{-1}\bigg]_{\mathrm{FF}}$ is the normalized scattered intensity of the form factor at low mAb concentration $c_0 = 2$ mg/ml obtained from measurements at the Diamond Light Source (see below).

Additional synchrotron SAXS measurements were performed on beamline B21 at Diamond Light Source, Didcot, UK. The incident X-rays had a wavelength of 0.09524 nm (13 keV), with a sample-to-detector (EigerX 4 M) distance of 3.69 m, corresponding to a $q$-range of $0.045-3.4$ nm$^{-1}$. Samples were loaded into the capillary using the BioSAXS sample robot. The temperature within the capillary and sample holder was set at T = 25 $^{\circ}$C. The continuously flowing samples were exposed for at least 10 frames (depending on initial sample volume and concentration), where each frame corresponds to an exposure of 1 second. Prior to averaging, sequential frames were investigated for inconsistencies caused, for example, by the presence of radiation damage. This was achieved by both visual inspections of the frames and by fitting the Guinier region for each individual frame. Before and after each sample measurement, identical measurements were performed on the buffer. The buffer frames were averaged and subtracted from the sample scattering. Calculation of S(q) followed essentially the same procedure as used for the in-house SAXS, with 2 mg/ml data used as the form factor. 
\newline

\noindent
\textbf{(ii) Dynamic and static light scattering.} Dynamic (DLS) and static (SLS) light scattering measurements were performed with a goniometer light scattering setup (3D LS Spectrometer, LS Instruments, AG), implementing a modulated 3D cross-correlation scheme to suppress  multiple scattering contributions \cite{urban1998characterization,block2010modulated}, and with an ALV/DLS/SLS-5022F, CGF-8F-based compact goniometer system (ALVGmbH, Langen, Germany). The light source for the 3D LS Spectrometer is a 660 nm Cobolt laser with a maximum power of 100 mW, while for the ALV instrument it  is a Helium-Neon laser operating at a wavelength $\lambda$ of 632.8 nm with maximum output power of 22 mW. All measurements on the 3D LS Spectrometer were performed at a scattering angle $\theta =$ 110$^{\circ}$, corresponding to a scattering vector  $q =  (4\pi n/\lambda) \sin(\theta/2) =$ 20.7  $\mu$m$^{-1}$, while those on the ALV instrument were performed at a scattering angle of $\theta =$ 104$^{\circ}$, corresponding to a scattering vector  $q =$ 20.8  $\mu$m$^{-1}$. Measurements were done at a temperature of 25 $^{\circ}$C.
Additional experiments were also performed on the 3D LS Spectrometer in a 2D geometry using pseudo-cross-correlation measurements at a scattering angle $\theta =$ 110$^{\circ}$ with the samples prepared for the SAXS experiments.

For DLS, intensity auto-correlation functions $g_{2}(q, \tilde{t})-1 $ vs. lag-time $\tilde{t}$ were analysed with a second-order cumulant function, using an iterative nonlinear fitting procedure \cite{frisken2001revisiting, mailer2015particle}:

\begin{equation}
    g_{2}(q, \tilde{t})-1 = B + \beta \Bigg\{ \mathrm{exp}{{(-\Gamma \tilde{t}}) \Big[ 1+\frac{1}{2}\mu_{2} \tilde{t}^{2} \Big] \Bigg\}^{2},}
\label{eq:1}
\end{equation}

\noindent where $B$ is the baseline, $\beta$ is the spatial coherence factor, $\Gamma$ is the relaxation rate (first cumulant) and $\mu_{2}$ is the second cumulant,  which characterizes deviations from the single exponential behavior. $\mu_{2}$ is related to the polydispersity of systems with $\sigma_* = \sqrt{\mu_{2}}/\Gamma$, where $\sigma_*$ is the normalized standard deviation of the size distribution. 
The apparent  hydrodynamic radius $R_\mathrm{h,app}$ of the scattered object was then calculated via the Stokes-Einstein relation:

\begin{equation}
R_\mathrm{h,app} = \frac{k_\mathrm{B}T }{6\pi\eta} \frac{q^{2}}{\Gamma},
\label{eq:2}
\end{equation}

\noindent where $\eta$ is the viscosity of the solvent and the term  $q^{2} /\Gamma = D_\mathrm{c}^{-1}$ is the inverse of the collective diffusion coefficient $D_\mathrm{c}$.\\

For SLS, we calculated the so-called excess Rayleigh ratio ($\Re_{\mathrm{ex}}$) from the measured scattering intensity \cite{Schurtenberger1991}. For samples with no multiple scattering contributions, \emph{i.e.}, negligible turbidity:

\begin{equation}
\Re_{\mathrm{ex}} =\frac{I(q)_{\mathrm{sample}}}{I(q)_{\mathrm{toluene}}} \Bigg(\frac{n_{\mathrm{sol}}}{n_{\mathrm{toluene}}}\Bigg)^{2}\Re_{\mathrm{toluene}},
\label{eq:3}
\end{equation}

\noindent where $I(q)_{\mathrm{sample}}$ and $I(q)_{\mathrm{toluene}}$ are the scattered intensities of the sample and the reference solvent toluene, respectively; $n_{\mathrm{sol}}$ and $n_{\mathrm{toluene}}$ are the refractive indexes for the solvent and toluene; $\Re_{\mathrm{toluene}}$ is the Rayleigh ratio for toluene in cm$^{-1}$. 
%For the 3D LS Spectrometer at $\lambda = 660$ nm and vertical/vertical polarized geometry (polarization of the incident and detected light) we have $\Re_{\mathrm{toluene}} = 0.8456 \times 10^{-5}$ cm$^{-1}$, while for the ALV instrument with $\lambda = 632$ nm and vertical/unpolarized geometry we have $\Re_{\mathrm{toluene}} = 1.364 \times 10^{-5}$ cm$^{-1}$, respectively, at $T = 25$ $^{\circ}$C~\cite{Sivokhin2021}.
For the 3D LS Spectrometer at $\lambda = 660$ nm and vertical/vertical polarized geometry (polarization of the incident and detected light) we have $\Re_{\mathrm{toluene}} = 0.8456 \times 10^{-5}$ cm$^{-1}$, for the measurements using 2D geometry on the 3D LS Spectrometer we applied a vertical/unpolarized geometry with $\Re_{\mathrm{toluene}} = 1.14 \times 10^{-5}$ cm$^{-1}$,  while for the ALV instrument with $\lambda = 632$ nm and vertical/unpolarized geometry we have $\Re_{\mathrm{toluene}} = 1.364 \times 10^{-5}$ cm$^{-1}$, respectively, at $T = 25$ $^{\circ}$C~\cite{Sivokhin2021}.

Finally, the apparent molecular weight of mAb ($M_\mathrm{w,app}$) as a function of concentration was then calculated using

\begin{equation}
M_\mathrm{w,app}  = \frac{\Re_{\mathrm{ex}}}{Kc}
\label{eq:5}
\end{equation}

\noindent where 

\begin{equation}
K = \frac{4\pi^2n^2_{\mathrm{sol}}\bigg(\frac{dn_{\mathrm{sam}}}{dc}\bigg)^2}{N_\mathrm{A} \lambda_0^4}.
\label{eq:6}
\end{equation}

\noindent Here, $c$ is the antibody concentration in mg mL$^{-1}$, the ratio  $\frac{dn_{\mathrm{sam}}}{dc}$ is  the refractive index increment of the antibody (= 0.194 mL mg$^{-1}$), $N_\mathrm{A}$ is the Avogadro number and $\lambda_0$ is the vacuum wavelength of the laser.

\subsection{Preparation of the single-molecule antibody model}
For determining the initial structure to be used in amino acid-level simulations, different models and levels of coarse-graining have been employed. The overall protocol consists of three steps. The same applies to both mAb-U and mAb-H. Further details for the former antibody can be found in Refs.~\cite{polimeni2024multi,gulotta2024combining}.
\newline
% \textbf{(i) Determination of the initial atomistic structure.} \fc{to be added}

\noindent\textbf{(i) Single-antibody Monte Carlo simulation at the amino acid level.} Starting from the atomistic antibody structure as determined in Section 4.1, we construct a coarse-grained model at the amino acid level~\cite{mahapatra2021self,polimeni2024multi}. In this representation, each amino acid is mapped to a spherical bead of diameter $\sigma_\mathrm{bead}^\mathrm{aa} = \left(6M_\mathrm{w}/\pi \rho\right)^{1/3}$, where $M_\mathrm{w}$ is the amino acid molecular weight (in g mol$^{-1}$) and $\rho = 1$ g mol$^{-1}$ Å$^{-3}$ is the average amino acid density~\cite{kaieda2014weak}. Each bead is placed at the center of mass of the atoms forming the corresponding amino acid. The resulting models consist of $\approx 1330$ beads. Their representation is given in Figure~\ref{fig:antibodies}a. We verified that using a model in which charges are slightly displaced as compared to the center of mass of the amino acids~\cite{vinterbladh2025intermolecular,cao2024coarse} does not qualitatively affect the results presented in the paper.

To determine the amino acids charge state, and consequently the net charge of the antibody, we perform Metropolis–Hastings Monte Carlo (MC) simulations using the \textsc{Faunus} simulation package~\cite{lund2008, stenqvist2013faunus}. Specifically, constant-pH MC simulations with titration moves are carried out on a single rigid antibody to equilibrate amino acid charges at the chosen pH and ionic strength. For both antibodies, the ionic strength is fixed at 7 mM and $\mathrm{pH}=6$, in line with experimental conditions. In \textsc{Faunus}, the charge titration move is implemented through a reactive MC scheme~\cite{johnson1994reactive}, which samples the protonation–deprotonation equilibrium of each amino acid by propagating back and forward the reaction $AH \rightleftarrows A + H$, where $AH$ and $A$ are, respectively, the amino acid protonated and deprotonated form. The equilibrium constants for each reaction $K_{\mathrm{a},i}$ are taken from literature~\cite{thurlkill2006pk} according to the amino acid species. The contribution of the titration move to the energy change in the MC scheme is given by:
\begin{eqnarray}
    \beta \Delta U = &-& \sum_{i} \ln \left[ \frac{N_{i}!}{(N_{i} +\nu)!}V^{\nu_{i}} \right] - \sum_{i} \ln\prod_i a_{i}^{\upsilon_{i}},
\label{eq:titration}    
\end{eqnarray}  
where, $N_i$ denotes the number of ionized amino acids of type $i$, $\nu_i$ is the stoichiometric coefficient (positive for products, negative for reactants), $V$ is the system volume, and $a_{i}$ is the activity of the amino acids of type $i$ (see Faunus documentation: https://faunus.readthedocs.io/). 
\newline

\noindent \textbf{(ii) Single-antibody Molecular Dynamics simulations at the atomistic level.} Using the protonation state of acidic and basic amino acids, i.e. glutamic acid, aspartic acid, lysine, arginine, and histidine, as assigned in (i), we perform all-atom molecular dynamics simulations on a single antibody in 7 mM NaCl aqueous solution using \textsc{gromacs} (versions 2020.3 and 2022.3)~\cite{abraham2015gromacs,markidis2015solving}. The antibody is described with the CHARMM36 force field~\cite{huang2017charmm36m}, while the TIP3P model is used for water~\cite{jorgensen1983comparison}. 
%The protonation state of acidic and basic amino acids, i.e. glutamic acid, aspartic acid, lysine, arginine, and histidine, is assigned based on each amino acid average charge as calculated from a one-protein MC simulation at pH 6 and 7mM NaCl.
For mAb-U, simulation details are provided in a previous work by some of us~\cite{polimeni2024multi}.

With reference to mAb-H, the antibody is inserted into a cubic simulation box of $17.77$ nm size, containing $177104$ water molecules, and the amount of Na cations and Cl anions to neutralize the system and reach the ionic strength of 7 mM. To equilibrate the system, we first perform an energy minimization by using the steepest descent algorithm with a tolerance of $1000$ kJ mol$^{-1}$ nm$^{-1}$. Then, temperature is equilibrated by performing $1$ ns simulation at $300$ K in the $NVT$ ensemble by using the velocity rescaling thermostat coupling algorithm with a time constant of 0.1 ps~\cite{bussi2007canonical}. Pressure is equilibrated by carrying out $1$ ns simulation at $1$ bar using the Parrinello-Rahman approach, with a time constant of $2$ ps.~\cite{parrinello1981polymorphic,nose1983constant}. After equilibration, a trajectory of $5.5$ $\mu$s is collected. The leapfrog integration algorithm~\cite{hockney1970potenitial} is used with a time step of $2$ fs. Periodic boundary conditions and minimum image convention are applied. The length of bonds involving H atoms is constrained using the LINCS procedure~\cite{hess1997lincs}. The cutoff of non-bonded interactions is set to $1$ nm, and electrostatic interactions are calculated using the smooth particle-mesh Ewald method.~\cite{essmann1995smooth}.
\newline

\noindent \textbf{(iii) Single-antibody Monte Carlo simulation at the amino acid level.} To assess whether conformational changes of the antibody during the atomistic simulation influence its charge state, we performed additional single-protein constant-pH MC simulations at the amino acid level, as in (i). Three representative configurations of the equilibrated antibody at 3000, 4000, and 5500 ns were coarse-grained to the amino acid level and used as input.
The results show only minor variations, with no substantial deviation from the initially calculated values, with the total net charge varying less than $0.5\%$. The amino acid representation of the antibody used throughout the paper is the one coarse-grained from the atomistic simulations at $5500$ ns for mAb-H and at $1500$ ns for mAb-U. 
For mAb-H, we also verified that using a different structural conformation of the antibody than the one chosen does not qualitatively impact the solution structure.

\subsection{Simulations of ensembles of antibodies at the amino acid level} 

With the chosen single-molecule configuration, we perform different types of MD simulations using an ensemble of antibodies at the amino acid level. In particular, we perform:

(i) MC simulations of $N_\mathrm{p}=20$ antibodies at different concentrations for both mAb-H and mAb-U. Antibodies are inserted in a simulation box of volume $V = N_\mathrm{p}M_\mathrm{w}/(c_\mathrm{p}N_\mathrm{A})$, 
%{\color{red} This expression is looks weird. What do you mean with the latter factor? Multiplication by 1e-27?} 
where $M_\mathrm{w}$ = 152 kDa is the antibody molecular weight for mAb-H and $M_\mathrm{w}$ = 148 kDa for mAb-U, $N_\mathrm{A}$ is Avogadro's number, and $c_\mathrm{p}$ is the antibody concentration. 
%expressed in mg/ml. 
The interaction potential between the \textit{i}-th and \textit{j}-th amino acid beads, $\Phi_\mathrm{aa}^\mathrm{Y}(r_{ij})$ is defined as,
\begin{equation} 
\beta \Phi_\mathrm{aa}^\mathrm{Y}(r_{ij})  = 
\ell_\mathrm{B} q_{i} q_{j}  \frac{e^{-r_{ij} /\ell_\mathrm{D}}}{r_{ij}}  + 
4\beta\epsilon_\mathrm{LJ}\left[\left( \frac{\sigma_{ij}}{r_{ij}} \right)^{12} - \left( \frac{\sigma_{ij}}{r_{ij}} \right)^{6} \right],
\label{eq:AA-Yuk}
\end{equation}
amounting to the sum of a screened electrostatic interaction (point-Yukawa) and a Lennard-Jones potential that accounts for van der Waals interactions. Here, $\sigma_{ij} = (\sigma^\mathrm{aa}_{\mathrm{bead},i} + \sigma^\mathrm{aa}_{\mathrm{bead},j})/2$ is the characteristic interaction distance~\cite{hagler1974energy} and the Debye length is calculated as
\begin{equation}
\ell_\mathrm{D}^{2} =  \left\{4 \pi \ell_\mathrm{B} \left[ \left(\frac{1}{1-\varphi}\right) Z_\mathrm{mAb}\rho_\mathrm{mAb} + 2\rho_\mathrm{buffer}\right]\right\}^{-1},
 \label{eq:Debye}
\end{equation}
where $\rho_\mathrm{buffer}$ is the number density of the dissociated buffer ions, while $\varphi$ is the excluded volume of a single antibody, calculated as the excluded volume of a hard sphere of diameter $\sigma_\mathrm{HS}=2R_\mathrm{g}$,
%equal to:
%\begin{equation}
%\phi_{HS} = \rho_{mAb} \frac{\pi \sigma_{HS}^{3}}{6},
% \label{eq:6}
%\end{equation}
%where $\rho_{mAb}$ is the mAb number density. For our simulations, we choose $\sigma_{HS}$ = $2R_{g}$, 
where $R_\mathrm{g}$ is the average radius of gyration of the antibody obtained from MD simulations, as in Ref.~\cite{polimeni2024multi}. 
%In particular, $R_{g}$ is equal to XXX for mAb-H and to 48.2 {\AA} for mAb-U.
The van der Waals attraction has an interaction strength $\epsilon_\mathrm{LJ}=0.075k_\mathrm{B}T$, following a previous work by some of us~\cite{polimeni2024multi}. Simulations are carried out for $10^4$ MC sweeps where, on each sweep, each antibody is attempted to be translated and rotated. Such simulations are performed with \textsc{faunus}. A table summarizing the most important parameters for this potential is given in the Supplementary Information.

(ii) MD simulations of $N=100$ mAb-H molecules where we employ the long-range bare Coulomb potential to account for the interaction between charged species.
The pair potential $\Phi_\mathrm{aa}^\mathrm{C}(r_{ij})$ for each bead belonging to the amino acid is defined as:
\begin{equation}
\Phi_\mathrm{aa}^\mathrm{C}(r_{ij}) = \frac{q_i q_j\sigma\epsilon_\mathrm{LJ}}{e^{*2}r_{ij}} + 
4\epsilon_\mathrm{LJ} \left[\left( \frac{\sigma}{r_{ij}} \right)^{12} - \left( \frac{\sigma}{r_{ij}} \right)^{6} \right],
\label{eq:AA-Coul}
\end{equation}
with $e^{*}=(4 \pi \varepsilon_0 \varepsilon_\mathrm{w} \sigma \epsilon_\mathrm{LJ})^{1/2}$ being the reduced unit of charge, $\varepsilon_0$ and $\varepsilon_\mathrm{w}$ the vacuum permittivity and the relative dielectric constant of water, respectively. Here, $\sigma$ corresponds to the diameter of each bead and is the unit of length. For converting to real units, we use $\sigma=5.74$ Å, corresponding to the average diameter of the amino acids belonging to the antibody. 
The particle-particle particle-mesh (PPPM) solver~\cite{hockney2021computer} is used for efficiently computing the long-range Coulomb interactions.
The addition of monovalent positive and negative ions ensures overall charge neutrality and incorporates the effect of buffer ions. However, reproducing the nominal salt concentration would require inserting an impractically large number of ions, rendering the simulations computationally prohibitive. We therefore fix the number of salt ions following the protocol of Ref.~\cite{camerin2025electrostatics}. 
As shown in Figure S4 by the radial distribution function of the center of mass of the antibodies, moderate variations in the number of added ions do not qualitatively alter the ensemble behavior of the antibodies.
Ions and counterions interact with a Weeks-Chandler-Anderson (WCA) potential and their diameter is always taken to be $0.1 \sigma$. Such simulations are performed with \textsc{lammps}~\cite{thompson2022lammps} in a cubic box with periodic boundary conditions in three dimensions in the NVT ensemble at
reduced temperature $T^*=k_\mathrm{B}T/\epsilon_\mathrm{LJ}=1.0$ employing a Langevin thermostat. Equilibration runs are carried out for at least $1\times10^6 \delta t$, with $\delta t=0.002\tau$ and $\tau=\sqrt{m \sigma^2/\epsilon_\mathrm{LJ}}$ the unit of time. A subsequent production run is carried out for at least $2\times10^7 \delta t$. We run simulations for three different replicas for each concentration.

\subsection{Measured quantities from simulation outcomes}

\textbf{(i) Static structure factor.} We calculate the effective structure factor $S_\mathrm{eff}(q)$ of the antibody ensemble as, 
\begin{equation}
S_\mathrm{eff}(q) = \frac{1}{F(q)} \frac{1}{N_\mathrm{tot}}
\left\langle \left[ \sum_{i=1}^{N_\mathrm{tot}}  \sin({\textbf{q}\cdot\textbf{r}_{i}})\right]^2 + \left[ \sum_{i=1}^{N_\mathrm{tot}}  \cos({\textbf{q}\cdot\textbf{r}_{j}})\right]^2 \right\rangle 
\end{equation}
where $N_\mathrm{tot}$ is the product of the number of antibodies and the number of amino acids, $F(q)$ is the form factor of the single molecule, and $r_{i}$ and $r_{j}$ identify the position of the \textit{$i^{th}$} and \textit\textit{$j^{th}$} amino acid.
\newline

\noindent\textbf{(ii) Electrostatic isopotential surfaces.} We visualize the charge distribution on the structure of the antibodies in two different ways. We employ the Adaptive Poisson-Boltzmann Solver (\textsc{apbs})~\cite{jurrus2018improvements} embedded in the Visual Molecular Dynamics (\textsc{vmd}) software~\cite{humphrey1996vmd} for having a global solution of the electrostatic potential within and around the molecule by solving the Poisson-Boltzmann partial differential equation. We also calculate the electrostatic potential of the two antibodies under investigation generated by treating the charges on each amino acid as if they were treated with Yukawa, as explained in the Results section. To this aim, we create a mesh grid around the antibody and for each point of the grid we determine the dimensionless electrostatic potential, given by Eq. \eqref{eq:ptYukpot}.
The visualization of the electrostatic isopotential surface is then performed using again \textsc{vmd}.

\subsection{Calculation of the effective interaction potential}

We calculate the effective interaction potential of mAb-H coarse-grained at the amino acid level by combining two complementary techniques, employing in both cases Yukawa interaction potentials. The use of implicit ions models allows for an efficient calculation of this descriptor. 
The first technique is the generalized Widom insertion method~\cite{mladek2011pair}. This is used at large distances $r\gtrsim 100 \text{ \AA}$ where it is known to yield better and more precise estimates than other methods. At shorter distances, instead, we rely on the Umbrella Sampling technique~\cite{blaak2015accurate,gnan2014casimir} by running multiple independent simulations at different (biased) equilibrium distances. 

We also calculate the rotationally averaged potential of mean force $U(r)$ of mAb-H by means of the Boltzmann inversion (BI) method~\cite{soper1996empirical}. This allows us to extract $U(r)$ from the radial distribution function of the center of mass of the antibodies as $U(r)=-k_\mathrm{B}T\ln g(r)$. To this aim, we run molecular dynamics simulations at $c=10$ mg/ml, which is the lowest limit for these computationally demanding runs, including counterions and explicit buffer ions.

\subsection{Calculation of osmotic compressibility and apparent hydrodynamic radii} 

We use liquid state theory to calculate the osmotic compressibility or $S(0)$ as a function of concentration~\cite{Goodstein1975,Naegele1996,cipelletti2025interacting}. The starting point is the link between the orientationally-averaged static structure factor and the pair distribution function $g(r)$ given by,

 \begin{equation}
	\label{pair-distribution}
	S(q) = 1 + 4 \pi \rho   \int_{0}^{\infty} r^2\left[g(r) - 1\right] \frac{\sin qr}{qr} \,dr\
\end{equation}

\noindent where $g(r)$ %or the local microstructure 
is calculated using a method based on the Ornstein-Zernike relation and the hybridized-mean spherical approximation (HMSA) closure relation first proposed by Zerah and Hansen \cite{Zerah1986}.   

The apparent hydrodynamic radius $R_\mathrm{h,app}$ and the interaction coefficient $k_\mathrm{D}$ for the collective diffusion coefficient are calculated using the relationship between the short time collective diffusion coefficient $D_\mathrm{c}^\mathrm{s}(q)$ and the ideal diffusion coefficient $D_0$ in the absence of interactions, given by \cite{Naegele1996, Banchio2008}
 \begin{align}
	\label{Dcoll}
	D_\mathrm{c}^\mathrm{s}(q) = D_0 \frac{H(q)}{S(q)}
\end{align}
where $H(q)$ is the hydrodynamic function that describes the effects of hydrodynamic interactions. $S(q)$ is calculated using HMSA as described above. For the calculation of $H(q)$ we assume pairwise additive hydrodynamic interactions, which should be accurate up to volume fractions of around $\varphi \leq 0.05$. For our antibodies this roughly corresponds to $c \leq 25$ mg/ml. We can thus write \cite{Neal1999}
\begin{align}
	\label{Hq}
	H(q) = 1 + 6 \pi \rho R_\mathrm{h}  \int_{0}^{\infty} r(g(r) - 1) F(qr) \,dr\ .
\end{align}
where $F(qr)$ is given by
\begin{align}
	\label{Fqr}
    F(qr) = \frac{\sin qr}{qr} +  \frac{\cos qr}{(qr)^2} -  \frac{\sin qr}{(qr)^3}.
    \end{align}

For small particles such as proteins, the measured diffusion coefficient corresponds to the so-called gradient diffusion coefficient given by 
\begin{align}
	\label{Dc}
	D_\mathrm{c} = \lim_{q \to 0} D_\mathrm{c}^\mathrm{s}(q) = D_0 \frac{H(0)}{S(0)} 
\end{align}
\noindent where $H(0)$ is related to the sedimentation velocity, $U_\mathrm{sed}$ \cite{Banchio2008}. 

This results in

\begin{align}
	\label{Rhapp}
	R_\mathrm{h,app} = R_{\mathrm{h},0} \frac{S(0)}{H(0)} 
\end{align}
\noindent where $R_{\mathrm{h},0}$ is the hydrodynamic radius of the mAb in the absence of interactions, i.e., at infinite dilution.

In the limit of low concentrations, i.e., in the second-virial regime, we write
\begin{align}
	\label{Hqkf}
	H(0) = \lim_{q \to 0}\left\{1 + 6 \pi \rho R_\mathrm{h} \int_{0}^{\infty}  r \left[e^{-\beta U(r)} - 1\right] F(qr) \,dr\right\} 
\end{align}
where we have again used $g(r) \approx$ exp$[-\beta U(r)]$ for sufficiently dilute antibody concentrations. We then rewrite Eq. \eqref{Hqkf} as
\begin{align}
	\label{H0approx}
	H(0) \approx 1 - k_\mathrm{f} c 
\end{align}
where the interaction coefficient $k_\mathrm{f}$ is given by
\begin{align}
	\label{eq:kf}
k_\mathrm{f} = 6 \pi R_\mathrm{h} \frac{N_\mathrm{A}}{M_\mathrm{w}} \int_{0}^{\infty}  r \left[1 - e^{-\beta U(r)}\right] F(qr) \,dr .
\end{align}
%{\color{red} JE: What is M?}
Using Eq. \eqref{eqn:kD} we can then calculate $k_\mathrm{D}$ from
\begin{align}
	\label{Rhapprox}
	\frac{R_\mathrm{h,app}}{R_{\mathrm{h},0}} = \frac{S(0)}{H(0)} \approx \frac{1-k_\mathrm{I}c}{1-k_\mathrm{f}c} \approx 1-(k_\mathrm{I}-k_\mathrm{f})c \approx 1 - k_\mathrm{D}c
\end{align}
where $k_\mathrm{D} = k_\mathrm{I} - k_\mathrm{f}$.

\normalsize
%\backmatter

\subsubsection*{Competing Interests}

The authors declare no financial or non-financial competing interests.

\subsubsection*{Acknowledgements}

We thank Gerardo Campos-Villalobos, Mikael Lund, and Jonathan S. Kingsbury and Charles G. Starr from Sanofi for useful discussions. This work was financially supported by Sanofi and the Swedish Research Council (VR; Grant No. 2016-03301, 2018-04627, 2019-06075 and 2022-03142). F.C. and P.S. acknowledge funding from the European Union’s Horizon Europe research and innovation programme under the grant agreement number 101106720, Marie Skłodowska-Curie Action Postdoctoral Fellowship, project many(Anti)Bodies - mAB. We benefited from access to the Diamond Light Source, Didcot, UK, where part of the SAXS measurements was performed at the beamline B21, and has been supported by iNEXT-Discovery, project number 871037, funded by the Horizon 2020 program of the European Commission, and we gratefully acknowledge the help of the local contacts Nathan Cowieson, Katsuaki Inoue, and Nikul Khunti. J.C.E. acknowledges funding from the National Science Centre, Poland, within the SONATA BIS grant no. 2023/50/E/ST3/00452. The computer simulations were also enabled by resources provided by the National Academic Infrastructure for Supercomputing in Sweden (NAISS) and the Swedish National Infrastructure for Computing (SNIC) at Lund University, partially funded by the Swedish Research Council through grant agreements no. 2022-06725 and no. 2018-05973. We also gratefully acknowledge the CINECA award under the ISCRA initiative, for the availability of high-performance computing resources and support. 
This work is part of the “LINXS Antibodies in Solution” research program and we acknowledge the financial support by the LINXS Institute of Advanced Neutron and X-ray Science.

\subsubsection*{Author Contributions}
Author contributions are defined based on CRediT (Contributor Roles Taxonomy). Conceptualization: F.C., E.Z., P.S.; Formal analysis: F.C., M.P., L.T., J.C.E., S.S., A.G., N.S-G.; Funding acquisition: F.C., A.S., P.S.; Investigation: F.C., M.P., L.T., J.C.E., A.S., E.Z., P.S.; Methodology: F.C., E.Z., P.S.; Project administration: F.C., A.S., P.S.; Software: F.C., M.P., L.T.; Supervision: A.S., E.Z., P.S.; Validation: F.C.; Visualization: F.C.; Writing – original draft: F.C., E.Z., P.S.; Writing – review and editing: F.C., M.P., L.T., J.C.E., S.S.,  A.S., E.Z., P.S..

\clearpage
\newpage
%\widetext
%\begin{widetext}
%\begin{center}
\onecolumngrid
\begin{center}
\large
\textbf{Beyond uniform screening: electrostatic heterogeneity dictates solution structure of complex macromolecules\\ \bigskip Supplementary Information}

\normalsize
\bigskip
Fabrizio Camerin\textsuperscript{ 1}, Marco Polimeni\textsuperscript{ 2}, Letizia Tavagnacco\textsuperscript{ 3, 4}, Jeffrey C.
Everts\textsuperscript{ 5, 6},\\Szilard Saringer\textsuperscript{ 1} Alessandro Gulotta\textsuperscript{ 1} Nicholar Skar-Gislinge\textsuperscript{ 1} Anna Stradner\textsuperscript{ 1, 7},\\ Emanuela Zaccarelli\textsuperscript{ 3, 4}, Peter Schurtenberger\textsuperscript{ 1, 7}\\
\medskip
\small
\textit{%
\textsuperscript{1}Division of Physical Chemistry, Department of Chemistry, Lund University, Lund, Sweden\\
\textsuperscript{2}Department of Pharmacy, Drug Delivery and Biophysics of Biopharmaceuticals, University of Copenhagen, Copenhagen, Denmark\\
\textsuperscript{3}CNR Institute of Complex Systems, Uos Sapienza, Piazzale Aldo Moro 2, 00185 Roma, Italy\\
\textsuperscript{4}Department of Physics, Sapienza University of Rome, Piazzale Aldo Moro 2, 00185 Roma, Italy
\textsuperscript{5}Institute of Theoretical Physics, Faculty of Physics, University of Warsaw, Poland\\
\textsuperscript{6}Institute of Physical Chemistry, Polish Academy of Sciences, Poland\\
\textsuperscript{7}LINXS Institute of Advanced Neutron and X-ray Science, Lund University, Lund, Sweden\\
}

\end{center}

\normalsize
\bigskip
%\twocolumngrid
%\renewcommand{\thefigure}{S\arabic{figure}}\setcounter{figure}{0}
\renewcommand{\theequation}{S\arabic{equation}}\setcounter{equation}{0}
\renewcommand{\thefigure}{S\arabic{figure}}\setcounter{figure}{0}
\renewcommand{\thetable}{S\arabic{table}}\setcounter{table}{0}

\section{Atomistic simulation snapshots}

\begin{figure}[h!]
\centering
\includegraphics[width=0.75\textwidth]{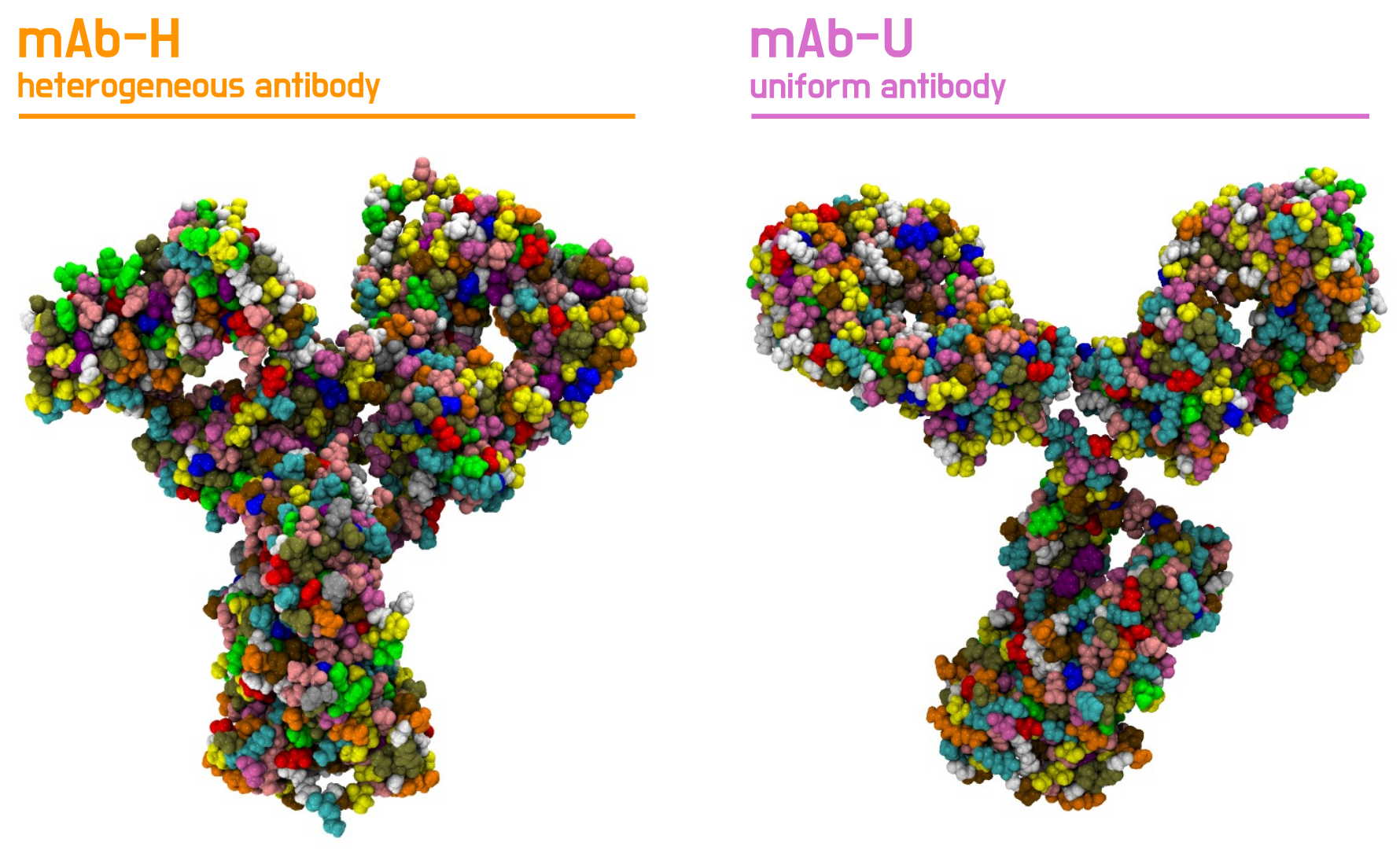}
\caption{\small \textbf{Atomistic simulation snapshot.} Representative simulation snapshots of the all-atom representation of (left) mAb-H and (right) mAb-U. Atoms are colored based on the amino acid to which they belong.}
\label{fig:models}
\end{figure}

\section{Simulation parameters}

\begin{table}[h!]
% \begin{tabular}{|c|c|c|c|}
% \hline
%             & single & $c=20$ mg/ml & $c=100$ mg/ml \\ \hline
% $\lambda_B$ &  &  &  \\ \hline
%  &  &  &  \\ \hline
%     &  &  &  \\ \hline
%   &  &  &  \\ \hline
% \end{tabular}
\begin{tabular}{|c|ccc|ccc|}
\hline
mAb & \multicolumn{3}{c|}{mAb-H}                            & \multicolumn{3}{c|}{mAb-U}                            \\ \hline
state & \multicolumn{1}{c|}{single} & \multicolumn{1}{c|}{$c=20$ mg/ml} & $c=100$ mg/ml & \multicolumn{1}{c|}{single} & \multicolumn{1}{c|}{$c=20$ mg/ml} & $c=100$ mg/ml \\ \hline
$\lambda_B$ [nm] & \multicolumn{1}{c|}{0.71} & \multicolumn{1}{c|}{0.71} & 0.71 & \multicolumn{1}{c|}{0.71} & \multicolumn{1}{c|}{0.71} & 0.71 \\ \hline
%$\lambda_D$ [nm]& \multicolumn{1}{c|}{} & \multicolumn{1}{c|}{} &  & \multicolumn{1}{c|}{} & \multicolumn{1}{c|}{} &  \\ \hline
$\kappa$ [nm$^{-1}$] & \multicolumn{1}{c|}{0.27} & \multicolumn{1}{c|}{0.31} & 0.43 & \multicolumn{1}{c|}{0.28} & \multicolumn{1}{c|}{0.32} & 0.47 \\ \hline
$\epsilon$ [$k_BT$] & \multicolumn{1}{c|}{0.075} & \multicolumn{1}{c|}{0.075} & 0.075 & \multicolumn{1}{c|}{0.075} & \multicolumn{1}{c|}{0.075} & 0.075  \\ \hline
\end{tabular}
\caption{\small \textbf{Simulation parameters.} Most relevant parameters for implicit ions Monte Carlo simulations at the amino acid level of mAb-H and mAb-U under 7 mM of ionic strength and different protein concentrations.}
\label{}
\end{table}

\clearpage
\newpage

\section{SLS and DLS data for the two antibodies}

\begin{figure*}[h!]
\centering
\includegraphics[width=\textwidth]{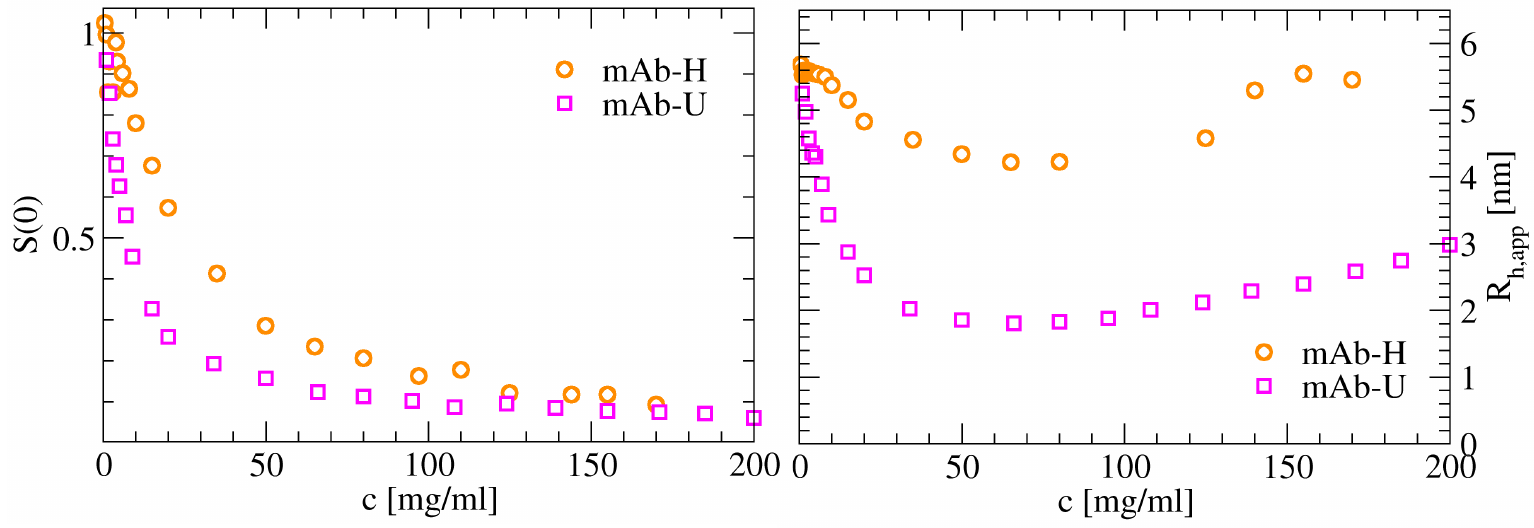}
\caption{\textbf{Experimental characterization.} (Left) Static structure factor at high length-scales $S(0)$ and (right) apparent hydrodynamics radius $R_{h,app}$ as a function of the antibody concentration $c$ for mAb-U and mAb-H.}
\label{fig:expscharact}
\end{figure*}

\section{SAXS data for the two antibodies}

\begin{figure}[h!]
\centering
\includegraphics[width=0.6\textwidth]{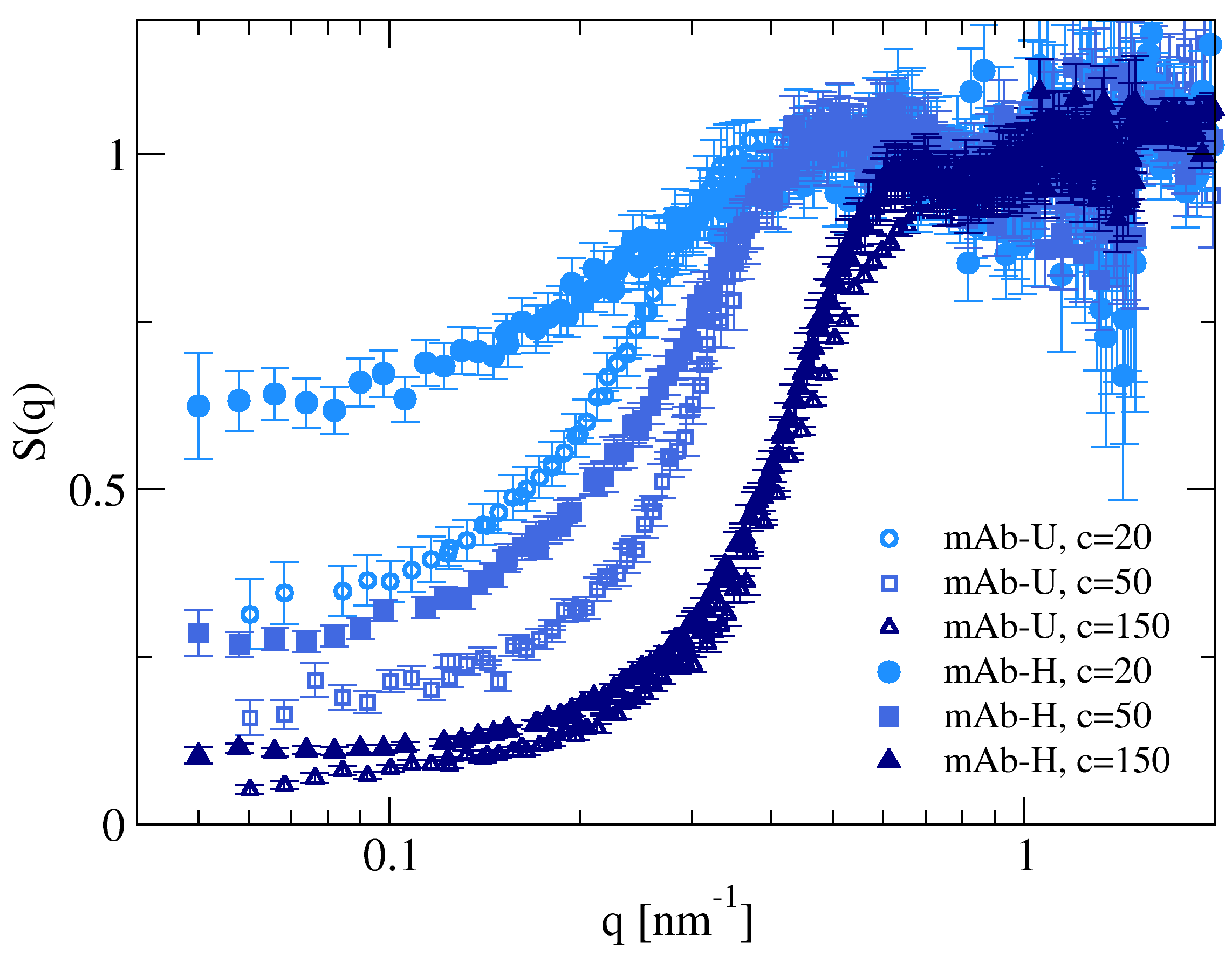}
\caption{\small \textbf{Experimental solution structure factors.} Static structure factors $S(q)$ as a function of the wavevector $q$ for mAb-U and mAb-H for three representative concentrations $c=20,50,150$ mg/ml.}
\label{fig:models}
\end{figure}

\section{Radial distribution function for explicit ions simulations}

\begin{figure}[h!]
\centering
\includegraphics[width=0.4\textwidth]{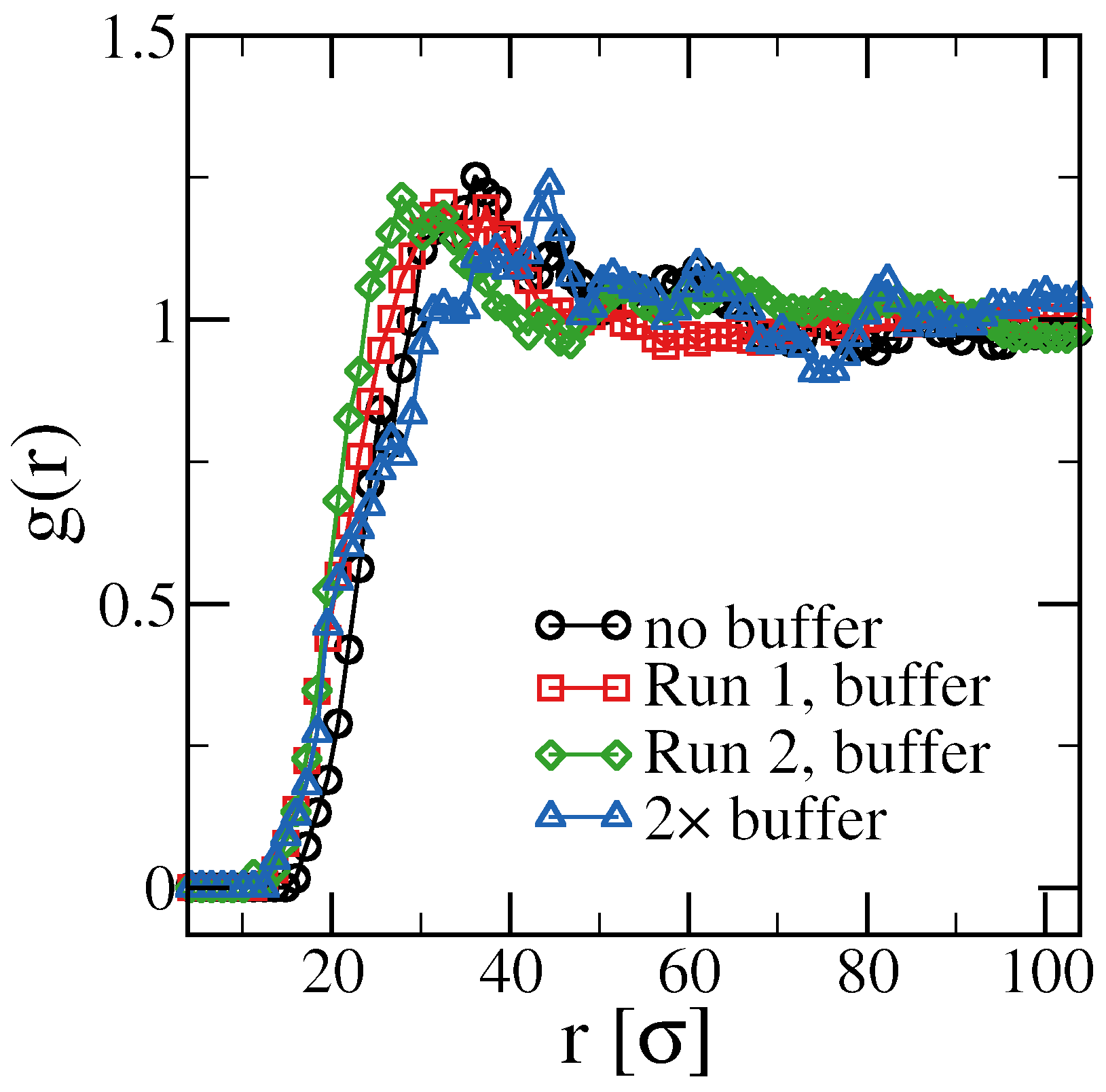}
\caption{\small \textbf{Effect of the buffer addition.} Center of mass radial distribution function $g(r)$ as a function of the distance $r$ for an ensemble of antibodies coarse-grained at the amino acid level at a concentration $c=10$ mg/ml for varying addition of the buffer ions. In particular, we show three cases: (i) no buffer is added, (ii) a buffer is added as explained in the main text for two different runs, and (iii) a buffer is added for twice that amount. No qualitative major difference is observed in the $g(r)$s.}
\label{fig:models}
\end{figure}

\twocolumngrid
\clearpage
\newpage

\bibliography{./main_arxiv.bbl}% common bib file

\end{document}